\documentclass{aa}  
\usepackage{graphicx}
\usepackage{txfonts}
\begin{document}
   \title{The Coma cluster magnetic field from Faraday rotation measures}
   %\subtitle{ }
   \author{A. Bonafede \inst{1,2}
     \and L. Feretti \inst{2}
     \and M. Murgia \inst{2,3}
     \and F. Govoni\inst{3}
     \and G. Giovannini \inst{1,2}
     \and D. Dallacasa,\inst{1,2}
     \and K. Dolag\inst{4}
     \and G. B. Taylor\inst{5} 
}

   \offprints{bonafede@ira.inaf.it} \institute{ Dip. di Astronomia,
     Univ. Bologna, via Ranzani 1, I-40120 Bologna, Italy \and INAF-
     Istituto di Radioastronomia, via Gobetti 101, I-40129, Italy
 \and INAF- Osservatorio Astronomico di Cagliari,Loc. Poggio
     dei Pini, Strada 54, I-09012, Capoterra (Ca) Italy  \and
     Max-Plank-Institut f{\"u}r Astrophysik, P.O. Box 1317, D-85741
     Garching, Germany \and Department of Physics and Astronomy,
     University of New Mexico 800 Yale Blvd NE, Albuquerque, NM 87131,
     USA, and Adjunct Astronomer at the National Radio Astronomy
     Observatory, USA}

   \date{Received ; accepted }
% 5 {} token are mandatory
  \abstract 
% context heading (optional) % {} leave it empty if  necessary 
{}%{To investigate the properties of the Intra Cluster Medium
  %magnetic field in the Coma Cluster} 
% aims heading (mandatory) 
{The aim of the present work is to constrain the Coma cluster
  magnetic field strength, its radial profile and power spectrum by
  comparing Faraday Rotation Measure (RM) images with numerical simulations
  of the magnetic field.}
% methods heading (mandatory) 
{We have analyzed polarization data for seven radio sources in the
  Coma cluster field observed with the Very Large Array at 3.6, 6 and
  20 cm, and derived Faraday Rotation Measures with kiloparsec scale
  resolution. Random three dimensional magnetic field models have been
  simulated for various values of the central intensity $B_0$ and
  radial power-law slope $\eta$, where $\eta$ indicates how the field
  scales with respect to the gas density profile.}
% results heading (mandatory) 
  {We derive the central magnetic field strength , and radial profile
    values that best reproduce the RM observations. We find that the
    magnetic field power spectrum is well represented by a Kolmogorov
    power spectrum with minimum scale $\sim$ 2 kpc and maximum scale
    $\sim$ 34 kpc. The central magnetic field strength and radial
    slope are constrained to be in the range
    ($B_0=$3.9 $\mu$G; $\eta=$0.4) and ($B_0=$5.4 $\mu$G; $\eta=$0.7)
    within 1$\sigma$. The best agreement between observations and
    simulations is achieved for $B_0=4.7 \mu$G;$\eta=$0.5. Values of
    $B_{0}>$7 $\mu$G and $<3$ $\mu$G as well as $\eta < 0.2$ and
    $\eta > 1.0$ are incompatible with RM data at 99\% confidence
    level. }
    % conclusions heading (optional), leave  it empty if necessary 
{} 
   \maketitle
%
%________________________________________________________________

\section{Introduction}
  It is now well established that the intracluster medium (ICM) of
  galaxy clusters is not only composed of thermal gas emitting in the
  X-ray energy band, but also of magnetic fields permeating the entire
  cluster volume (see Ferrari et al., 2008 for a recent review). This
  is directly demonstrated by the detection of large, diffuse
  synchrotron radio sources such as radio halos and radio relics, in an
  increasing number of galaxy clusters (see e. g. Venturi et al. 2008,
  Giovannini et al. 2009). In these clusters it is possible to
  estimate the average ICM magnetic field under the minimum energy
  hypothesis (which is very close to equipartition conditions) 
  or by studying the Inverse
  Compton hard X-ray emission (e.g. Fusco Femiano et al. 2004).\\ The
  ICM magnetic field is also revealed by the analysis of polarized
  emission of radio sources located at different projected distances
  with respect to the cluster center. The interaction of the ICM, a
  magneto-ionic medium, with the linearly polarized synchrotron
  emission results in a rotation of the wave polarization plane
  (Faraday Rotation), so that the observed polarization angle,
  $\Psi_{obs}$ at a wavelength $\lambda$ differs from the intrinsic
  one, $\Psi_{int}$ according to:
\begin{equation}
\label{eq:psiRM}
 \Psi_{obs}(\lambda) = \Psi_{int}+\lambda^2 \times RM,
\end{equation}
where RM is the Faraday Rotation Measure. This is related to the
magnetic field component along the line-of-sight ($B_{//}$)
weighted by the thermal gas density ($n_e$) according to:
\begin{equation}
\label{eq:RM}
RM \propto \int_{los}n_e(l) B_{//}(l) dl.
\end{equation}
Therefore, once the thermal gas density distribution is inferred from
X-ray observations, RM studies give an additional set of information
about the cluster magnetic field. This is the only way, so far, to
study the intracluster magnetic field in clusters where diffuse radio
emission sources are not directly observed.\\ The Coma cluster
magnetic field has been studied in the past using all three approaches
mentioned above. The first investigation of the magnetic field was
performed by Kim et al. (1990).  They analyzed 18 bright radio-sources
in the Coma cluster region, obtaining RM maps at $\sim 20 ''$ ($\sim$
9.2 kpc) resolution and found a significant enhancement of the RM in
the inner parts of the cluster. Assuming a simple model for the
magnetic field reversal length, they derived a field strength of
$\sim$ 2 $\,mu$G. A complimentary study was performed by Feretti et
al. (1995) studying the polarization properties of the extended radio
galaxy NGC 4869. From the average value of RM and its dispersion
across the source, they deduced a magnetic field of $\sim$6 $\mu$G
tangled on scales of $\sim$ 1 kpc, in addition to a weaker magnetic
field component of $\sim$0.2 $\mu$G , uniform on a cluster core radius
scale.\\ From the Coma radio halo, assuming equipartition, a magnetic
field estimate of $\sim 0.7-1.9$ $\mu$G is derived (Thierbach et
al. 2003), while from the Inverse Compton hard X-ray emission a value
of $\sim$0.2 $\mu$G has been derived by Fusco Femiano et al. (2004),
although new hard X-ray observations performed with a new generation
of satellites did not find such evidence of non-thermal emission (Wik
et al. 2009 using XMM and Suzaku data, Lutovinov et al. 2008 using
ROSAT, RXTE and INTEGRAL data, Ajello et al. 2009 using XMM-Newton,
Swift/XRT, Chandra and BAT data; see Sec. \ref{sec:comparison}).
\\ However, the discrepancy between these values is not surprising:
equipartition and IC estimates, in fact, rely on several assumptions,
and are cluster volume averaged estimates, while the RM is sensitive
to the local structures of both the thermal plasma and the cluster
magnetic field component that is parallel to the line of
sight. Furthermore, the equipartition estimate should be used with
caution, given the number of underlying assumptions. For example, it
depends on the poorly known particle energy distribution, and in
particular on the low energy cut-off of the emitting electrons (see e. g. Beck \& Krause 2005).\\ These different estimates provide
direct evidence of the complex structure of the cluster magnetic field
and show that magnetic field models where both small and large scale
structure coexist must be considered. The intracluster magnetic field
power spectrum has been investigated in several works (Vogt \& Ensslin
2003, 2005; Murgia et al., 2004; Govoni et al. 2006, Guidetti et
al. 2008, Laing et al. 2008). In addition, a radial decline of the
magnetic field strength is expected from magneto-hydrodynamical
simulations performed with different codes (Dolag et al. 1999, 2005;
Brueggen et al. 2005, Dubois et al. 2008, Collins et al. 2009, Dolag
\& Stasyszyn 2008, Donnert et al. 2009a), as a result of the
compression of thermal plasma during the cluster gravitational
collapse.\\ In this paper we present a new study of the Coma cluster
magnetic field.  We analyzed new Very Large Array (VLA) polarimetric
observations of seven sources in the Coma cluster field, and used the
FARADAY code developed by Murgia et al. (2004) to derive the magnetic
field model that best represents our data through numerical
simulations.\\ The Coma cluster is an important target for a detailed
study of cluster magnetic fields. It is a nearby cluster (z=0.023), it
hosts large scale radio emission (radio halo, radio relic, bridge) and
a wealth of data are available at different energy bands, from radio
to hard X-rays.\\ The paper is organized as follows: in
Sec. \ref{sec:X} the properties of the X-ray emitting gas are
described, in Sec. \ref{sec:Obs} radio observations are presented, and
the radio properties of the sources are analyzed. RM analysis is
presented in Sec.  \ref{sec:RMobs} while in Sec.\ref{sec:Sim_setup}
the adopted magnetic field model is described. The method we adopted
to compare simulations and observations is reported in
Sec. \ref{sec:errors}, numerical simulations are presented in
Sec. \ref{sec:Sim} and results are discussed in
Sec. \ref{sec:comparison}. Finally, conclusions are reported in
Sec. \ref{sec:concl}. \\ A $\Lambda$CDM cosmological model is assumed
throughout the paper, with $H_0=71$ km/s/Mpc, $\Omega_M$=0.27,
$\Omega_{\Lambda}$=0.73. This means that 1 arcsec corresponds to 0.460
kpc at z=0.023.\\

%__________________________________________________________________
\section{Thermal component from X-ray observations}
\label{sec:X}
\begin{figure*}[ht]
\centering
\includegraphics[width=\textwidth]{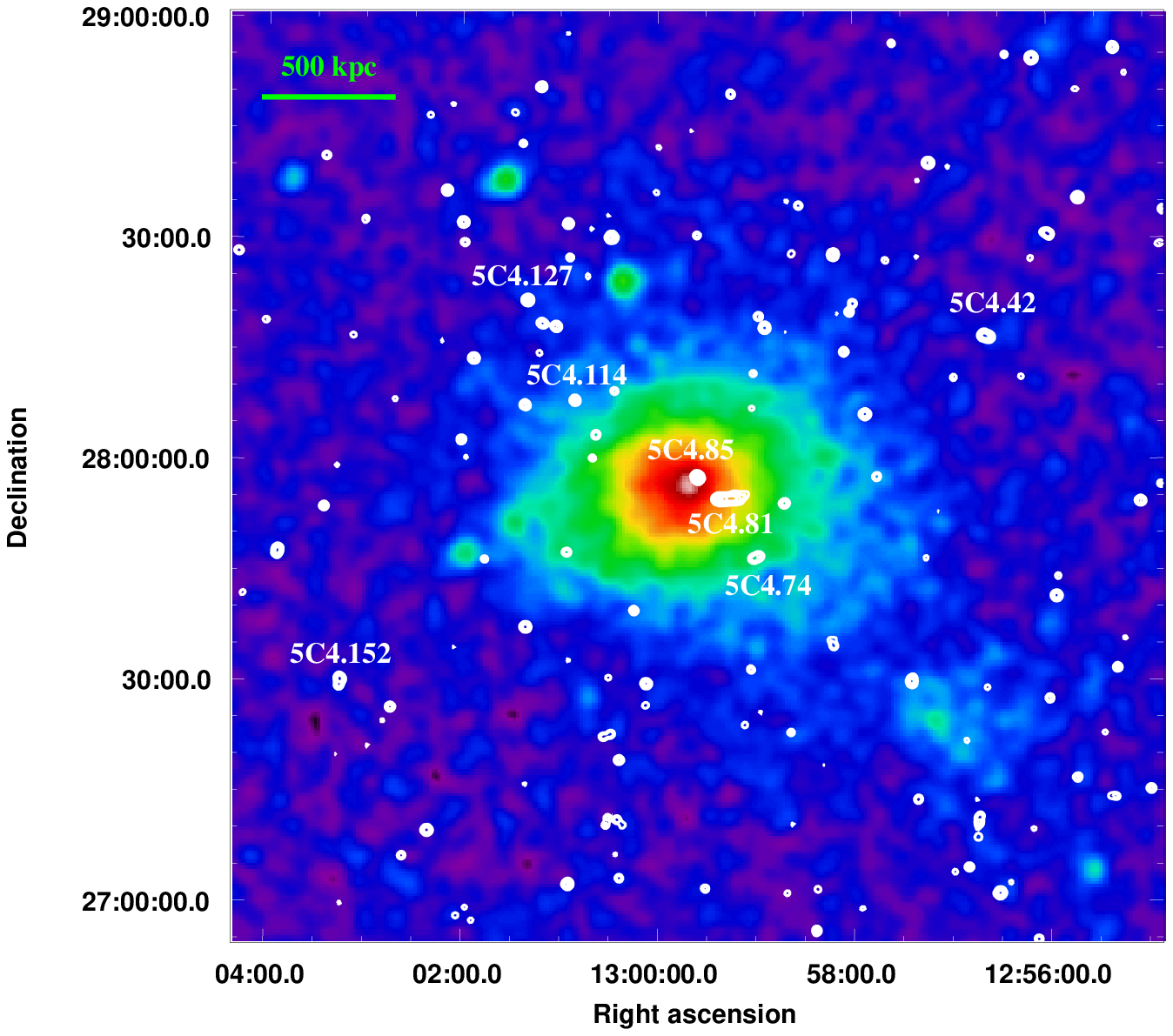}
\caption{Colors: Coma X-ray emission from the ROSAT All Sky Survey in
  the energy band [0.1, 2.4] kev. Contours: Coma radio emission at 1.4
  GHz from the NVSS. The beam FWHM is 45$''\times$45$''$, contours
  start from 1.5 mJy/beam and are spaced by a factor of 2. The
  observed sources are labelled.}
\label{fig:comaX}
\end{figure*}

The study of the magnetic field through the Faraday RM requires
knowledge of the properties of the thermal gas (see
Eq. \ref{eq:RM}). This information can be derived from X-ray
observations. In Fig. \ref{fig:comaX} the X-ray emission of the Coma
cluster is shown in colors. X-ray observations in the energy band
$0.1-2.4$ keV have been retrieved from the ROSAT All Sky Survey data
archive.  After background subtraction the image has been divided by
the exposure map and smoothed with a Gaussian of $\sigma=60''$.\\ The
radio contours of the NVSS (NRAO VLA Sky Survey) at 1.4 GHz are
overlaid onto the X-ray emission in Fig. \ref{fig:comaX}. The location
of the sources studied in this paper are marked with their names.
Note that the extended radio emission of the radio halo, relic and
bridge are completely resolved out in the NVSS image due to the lack
of extremely short baselines. \\ The X-ray emission is from thermal
bremsstrahlung, and can be used to trace the thermal particle
distribution in the ICM. The distribution of the gas is well
reproduced by the so-called $``\beta$-model'' (Cavaliere \& Fusco
Femiano, 1976), reported in Eq. \ref{eq:betaModel}:
\begin{equation}
n_e(r)=n_0 \left(1+\frac{r^2}{r^2_c}\right)^{-\frac{3}{2}\beta},
\label{eq:betaModel}
\end{equation}
where $r$ indicates the radial distance from the cluster center, $r_c$
is the cluster core radius, $n_0$ is the cluster central gas
density. The cluster center is RA $=$ 12h 59m 41.5s; DEC $=$
27$^{\circ}$ 56$'$ 20$''$.  We use the values derived by Briel et
al. (1992), corrected for the cosmology adopted in this paper. They
found :
\begin{itemize}
\item $\beta=$0.75$\pm$0.03;
\item $r_c=$291$\pm$17 kpc;
\item $n_0=$3.44$\pm$0.04 10$^{-3}$ $cm^{-3}$.
\end{itemize}

%--------------------------------------------------------------------------
\section{Radio observations and images}
\label{sec:Obs} 
\subsection{VLA observations and data reduction}
We selected from NVSS a sample of sources having a peak flux density
larger than 45 mJy, located in a radius of $1^{o}$ from the cluster
X-ray center center ($\simeq 5 r_{c}$), and which have indication of
polarization from Kim et al. (1990). A further selection was performed
on the basis of the position of the sources. Observations have been performed at the
VLA on 7 sources: 5C4.85 (NGC 4874), 5C4.81 (NGC 4869), 5C4.74,
5C4.114, 5C4.127, 5C4.42, and 5C4.152.  Radio observations were
performed using the 6 cm and 3.6 cm bands for all of the source except
5C4.114. This source being weaker, was observed at 20 cm and 6 cm. 
The sources were observed at two frequencies within each band,
in order to have 4 frequency observations: 4.535 GHz, 4.935 GHz, 8.085
GHz, and 8.465 GHz. In addition 4.735 GHz observations were performed on
sources 5C4.85 and 5C4.74. Due to technical issues the observing time
for the source 5C4.152 was reduced, and it was necessary to increase
the signal-to-noise ratio in the 3.6 cm band.  To increase the
signal-to-noise ratio, the data at 8.085 and 8.465 GHz were averaged
together and a single image at 8.275 GHz was obtained.  For the source
5C4.114 we have observations at 1.365 GHz, 1.516 GHz, 4.535 GHz and
4.935 GHz. Details of the observations are reported in Table
\ref{tab:radioobs}. The source 3C286 was used as both primary flux
density calibrator\footnote{we refer to the flux density scale by
  Baars et al. (1990)} and as absolute reference for the electric
vector polarization angle. The source 1310+323 was observed as both
a phase
and parallactic angle calibrator. \\ We performed standard calibration
and imaging using the NRAO Astronomical Imaging Processing Systems
(AIPS). Cycles of phase self-calibration were performed to refine
antenna phase solutions on target sources, followed by a final
amplitude and gain self-calibration cycle in order to remove minor
residual gain variations. Total intensity, I, and Stokes parameter Q
and U images have been obtained for each frequency separately. After
cleaning, radio images were restored with a naturally weighted beam.
The final images were then convolved with a Gaussian
beam having FWHM $=$ 1.5$''\times$1.5$''$ ($\sim$ 0.7$\times$0.7
kpc). Polarization intensity $P=\sqrt{U^2+Q^2}$, Polarization angle
$\Psi=\frac{1}{2}atan(U,Q)$ and fractional polarization
$FPOL=\frac{P}{I}$ images were obtained from the I, Q and U
images. Polarization intensity images have been corrected for a 
positive bias. The calibration errors on the measured fluxes are
$\sim$ 5\%. \\
\begin{table*} 
\caption{VLA observations of radio galaxies in the Coma cluster field.}          
\label{tab:radioobs}      
\centering          
\begin{tabular}{|c c c c c c c c|}     % 8 columns 
\hline\hline       
Source       & RA           &  DEC      &  $\nu$    & Bandwidth & Config. & Date & Time on Source   \\
             & (J2000)      & (J2000)   &   (GHz)   & (MHz)     &         &      & (Hours)  \\\hline                    
5C4.85 &  12 59 35.3  & +27 57 36 &  8.085 - 8.465  & 50  & B  & Jul 06         & 2.6   \\ 
       &              &           &  4.535 - 4.935  & 50  & B  & Oct 07         & 2.6   \\
       &              &           &      4.735      & 50  & B  & May 09         & 2.6   \\
       &              &           &                 &     &    &                &  \\
5C4.81 &  12 59 22.8  & +27 54 40 &  8.085 - 8.465  & 50  & B  & Jul 06         & 2.6   \\
       &              &           &  4.535 - 4.935  & 50  & B  & Jul 06         & 2.7   \\
       &              &           &      8.465      & 50  & C  & Nov 90         & 3.8   \\
       &              &           &         &       &     &               &       \\
5C4.74 &  12 58 59.4  & +27 46 46 &  8.085 - 8.465  & 50  & B  & Jul 06         & 2.7   \\
       &               &          &  8.085 - 8.465  & 50  & C  & Apr 08         & 5.4  \\
       &               &          &  4.535 - 4.935  & 50  & B  & Jul 06         & 2.7  \\
       &              &           &      4.735      & 50  & B  & May 09         & 2.6   \\
       &              &           &                 &     &    &                &       \\
5C4.114&  13 00 50.6  & +28 08 03 &  1.365 - 1.515  & 25  & A  & Dec 08         & 4.6   \\
       &              &           &  4.535 - 4.935  & 50  & B  & Apr 09         & 5.1   \\
       &              &           &         &     &     &               &       \\
5C4.127&  13 01 20.1  & +28 21 38 &  8.085 - 8.465  & 50  & B  & Jul 06         & 2.6   \\
       &              &           &  4.535 - 4.935  & 50  & B  & Oct 07         & 2.9   \\
       &                    &           &         &     &     &               &       \\
5C4.42 &  12 56 38.2  & +28 16 43 &  8.085 - 8.465  & 50  & B  & Jul 06         & 2.5   \\
       &              &           &  4.535 - 4.935  & 50  & B  & Jul 06         & 2.6   \\
 &              &           &         &     &     &               &       \\
5C4.152&  13 03 14.4  & +27 30 06 &  8.085 - 8.465  & 50  & B  & Aug 06         & 1.2   \\
       &              &           &  4.535 - 4.935  & 50  & B  & Aug 06         & 1.5   \\
\hline                  
\multicolumn{8}{l}{\scriptsize Col. 1: Source name; Col. 2, Col. 3: Pointing position (RA, DEC);
Col. 4: Observing frequency;}\\
\multicolumn{8}{l}{\scriptsize Col 5: Observing bandwidth; Col. 6: VLA configuration; 
Col. 7: Dates of observation;}\\
\multicolumn{8}{l}{\scriptsize Col. 8: Time on source (flags taken into account).}\\ 
\end{tabular}
\end{table*}
\begin{table*}
\caption{ Total and polarization intensity radio images. Images are restored with a beam of 1.5$''\times$1.5$''$ }
\label{tab:radiomaps}
\centering
\begin{tabular} {|c c c c c c c c|} 
\hline\hline
Source name & $\nu $  &   $\sigma$(I) & $\sigma$(Q) & $\sigma$(U) & Peak brightness  &Flux density & Pol. flux \\
             & (GHz)          & (mJy/beam)   & (mJy/beam)  & (mJy/beam)  & (mJy/beam)         & (mJy)       & (mJy)     \\ 
\hline
5C4.85  &  4.535    & 0.018 & 0.015 & 0.015  &  5.8 &  75.7   & 7.5  \\
        &  4.735   & 0.017 & 0.016 & 0.016  &  5.7 &  75.6   & 7.6   \\
        &  4.935   & 0.019 & 0.017 & 0.017  &  5.8 &  75.2   & 7.5  \\
        &  8.085   & 0.032 & 0.021 & 0.023  &  3.3 &  38.6   & 4.2    \\
        &  8.465   & 0.034 & 0.022 & 0.022  &  3.1 &  36.4   & 4.0\\
\hline
5C4.81  &  4.535 & 0.025 & 0.022 & 0.024  &  2.7 &  66.7   & 12.1 \\ 
        &  4.935 & 0.025 & 0.022 & 0.022  &  2.7 &  58.2   & 10.5  \\
        &  8.085 & 0.022 & 0.019 & 0.021  &  2.4 &  25.2   & 5.5  \\
        &  8.465 & 0.015 & 0.014 & 0.014  &  2.3 &  54.0*   &  11.3  \\
\hline
5C4.74  &  4.535 & 0.025 & 0.023 & 0.023  &  2.7 &  17.0  & 3.4 \\ 
        &  4.735 & 0.019 & 0.016 & 0.016  &  2.3 &  13.4  & 2.7\\
        &  4.935 & 0.023 & 0.021 & 0.021  &  2.6 &  15.6  & 3.1\\
        &  8.085 & 0.015 & 0.013 & 0.013  &  1.7 &   9.1  & 2.2\\
        &  8.465 & 0.014 & 0.013 & 0.013  &  1.6 &   8.6  & 2.1  \\ 
\hline
5C4.114 &  1.365 & 0.040 & 0.027 & 0.027  &  11.3 & 47.0& 5.9\\
        &  1.516 & 0.034 & 0.021 & 0.022  &  10.4 & 42.6 &5.5\\
        &  4.535 & 0.014 & 0.013 & 0.013  &  4.2 & 16.4 &3.0\\
        &  4.935 & 0.014 & 0.013& 0.013   &  3.8 & 14.9 &2.8\\
\hline
5C4.127 &  4.535 & 0.028 & 0.019 & 0.017  & 56.3 & 72.5   & 3.9  \\ 
        &  4.935 & 0.024 & 0.020 & 0.017  & 56.4 & 72.9  & 3.9\\
        &  8.085 & 0.023 & 0.021 & 0.022  & 52.2 & 62.3  & 3.4\\
        &  8.465 & 0.025 & 0.021 & 0.023  & 51.6 & 61.4   & 3.3  \\ 
\hline
5C4.42  &  4.535 & 0.023 & 0.022 & 0.022  & 7.2  &  63.2 & 8.0 \\
        &  4.935 & 0.023 & 0.022 & 0.022  & 6.5  &  57.8 & 7.3 \\
        &  8.085 & 0.022 & 0.020 & 0.020  & 3.8  &  33.2  & 4.1 \\
        &  8.465 & 0.021 & 0.020 & 0.019  & 3.6  &  31.0  & 4.0   \\
\hline
5C4.152 &  4.535 & 0.026 & 0.024 & 0.025  &  7.0 &  22.7  & 2.9  \\
        &  4.935 & 0.026 & 0.025 & 0.025  &  6.4 &  20.3  & 2.4 \\ 
        &  8.275 & 0.021 & 0.019 & 0.019  &  4.1 &  12.3  & 1.8\\
\hline \multicolumn{8}{l}{\scriptsize Col. 1: Source name;
  Col. 2: Observation frequency; Col. 3, 4, 5:
  RMS noise of the I, Q, U images; }\\ \multicolumn{8}{l}{\scriptsize
  Col. 7: Peak brightness; Col. 8: Flux density; Col. 9: Polarized
  flux density.}\\ \multicolumn{8}{l}{\scriptsize *The higher
    flux measured at 8.465 GHz is derived by combining together B and
    C array observations. }
\end{tabular}
\end{table*}

\begin{figure*}[htb]
\centering
\includegraphics[width=0.9\textwidth]{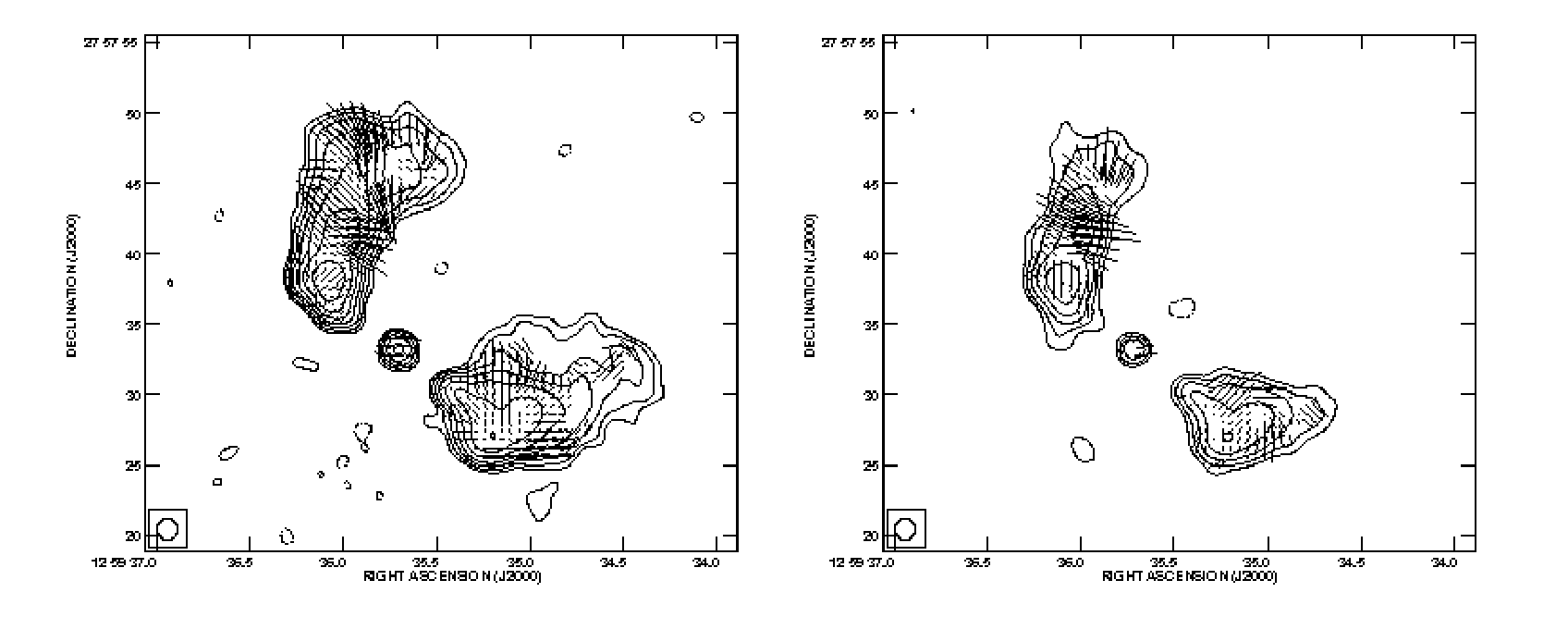}
\caption{Source 5C4.85. Total intensity radio contours and
  polarization vectors at 4.535 GHz (left) and 8.465 GHz (right). The
  bottom contour corresponds to a 3$\sigma$ noise level, contours are
  then spaced by a factor of 2. E vectors are superimposed: 
  the orientation indicates the direction of the E field, while the line
  length is proportional to the fractional polarization intensity
  (1$''$ corresponding to 10\%).}
\label{fig:85_pol}
\end{figure*}

\begin{figure*}[htb]
\centering\includegraphics[width=0.9\textwidth]{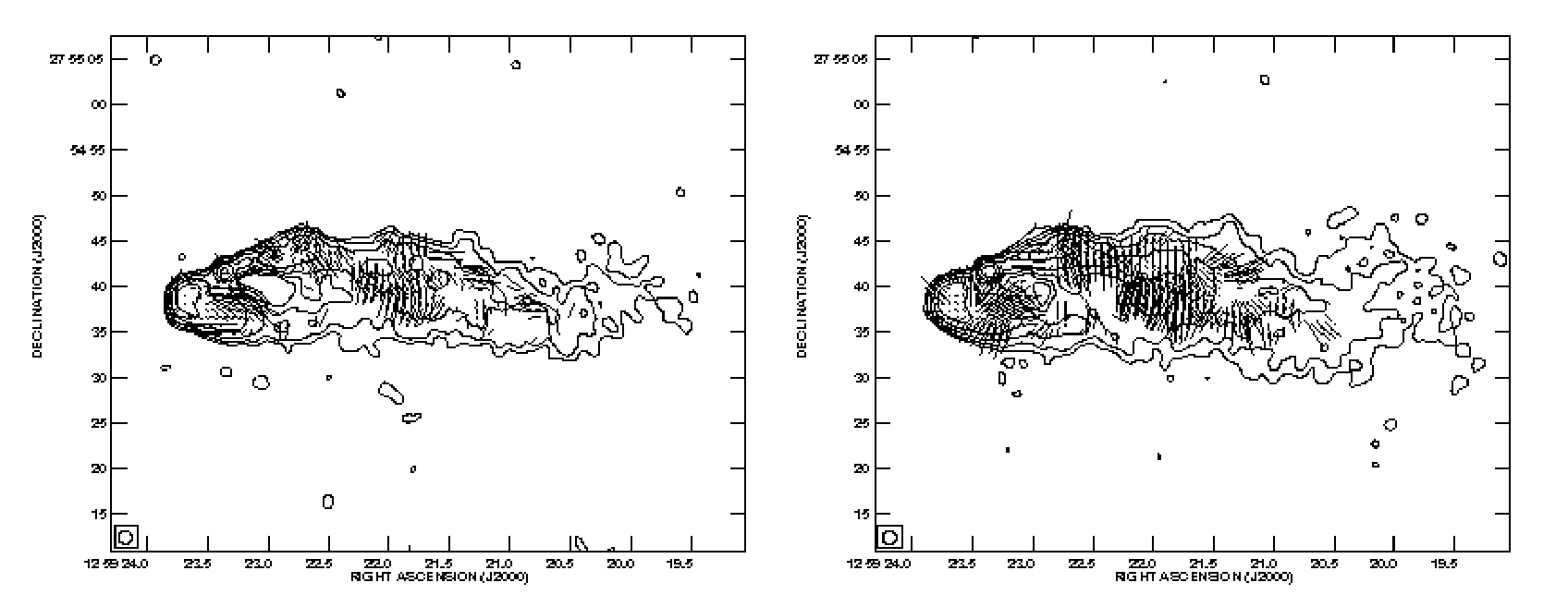}
\caption{Source 5C4.81. Total intensity radio
contours and polarization vectors at 4.535 GHz (left) and 8.465 GHz
(right). The bottom contour corresponds to a 3$\sigma$ noise level, and 
contours are then spaced by a factor of 2. E vectors are superimposed:
orientation indicates the direction of the E field, while line
length is proportional to the fractional polarization intensity (1$''$
corresponding to 10\%). }
\label{fig:81_pol}
\end{figure*}

\begin{figure*}[htb]
\centering\includegraphics[width=0.9\textwidth]{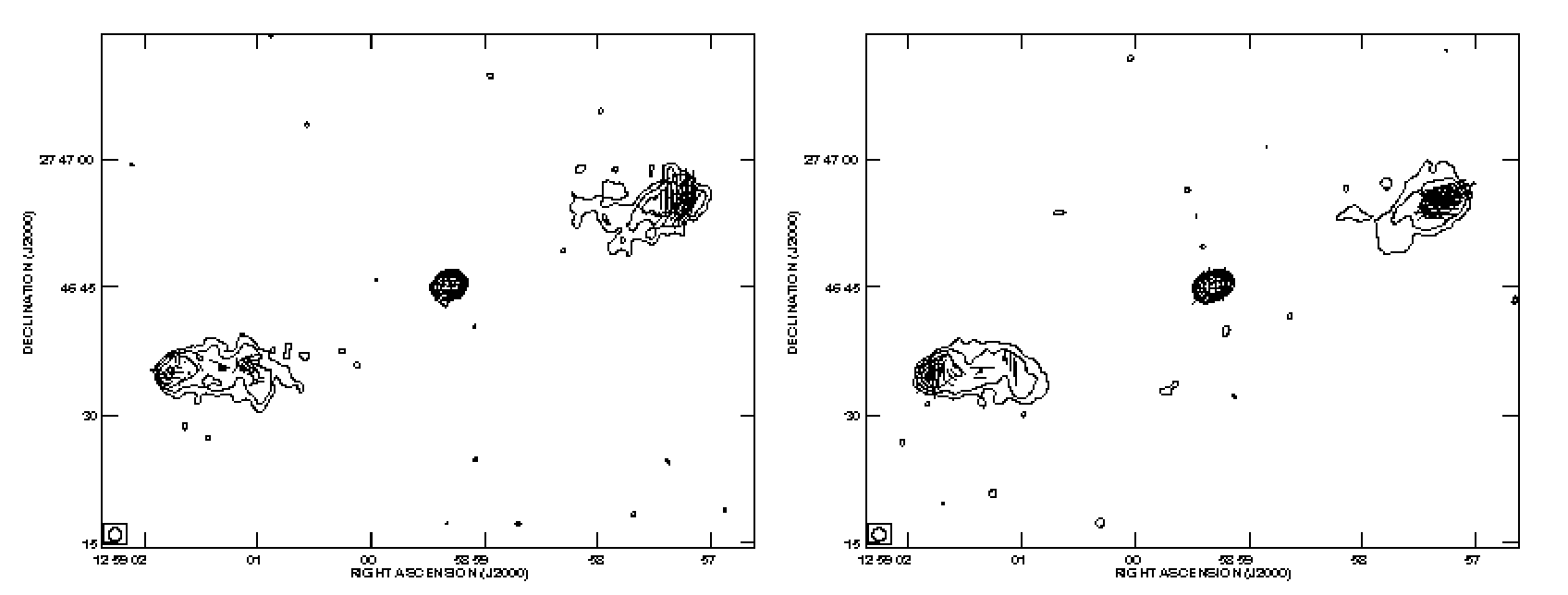}
\caption{Source 5C4.74. Total intensity radio
contours and polarization vectors at 4.535 GHz (left) and 8.465 GHz
(right). The
  bottom contour corresponds to a 3$\sigma$ noise level, contours are
  then spaced by a factor of 2. E vectors are superimposed: 
  the orientation indicates the direction of the E field, while the line
  length is proportional to the fractional polarization intensity
  (1$''$ corresponding to 10\%).  }
\label{fig:74_pol}
\end{figure*}

\begin{figure*}[htb]
\centering\includegraphics[width=0.9\textwidth]{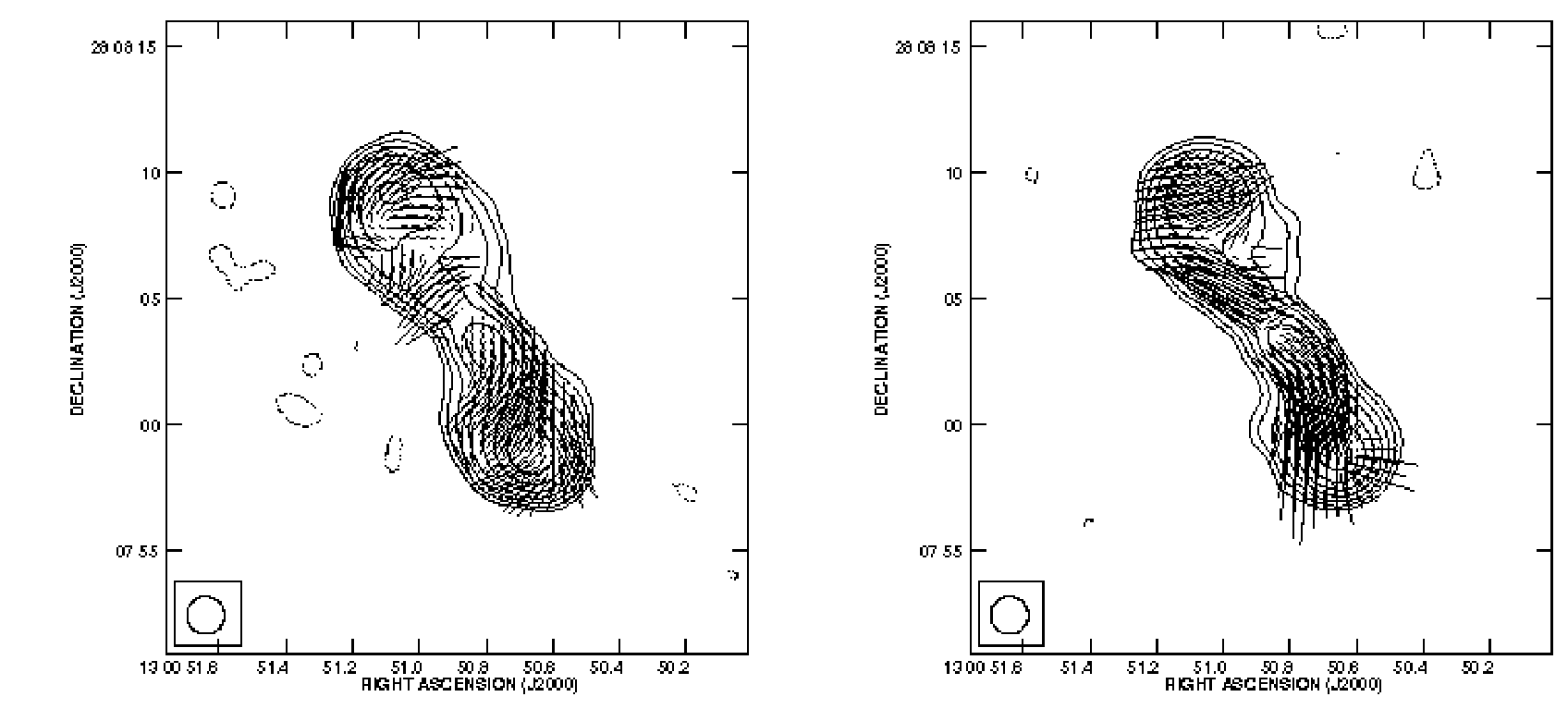}
\caption{Source 5C4.114. Total intensity radio
contours and polarization vectors at 1.365 GHz (left) and 4.935 GHz
(right). The
  bottom contour corresponds to a 3$\sigma$ noise level, contours are
  then spaced by a factor of 2. E vectors are superimposed: 
  the orientation indicates the direction of the E field, while the line
  length is proportional to the fractional polarization intensity
  (1$''$ corresponding to 10\%).  }
\label{fig:114_pol}
\end{figure*}

\begin{figure*}[htb]
\centering\includegraphics[width=0.9\textwidth]{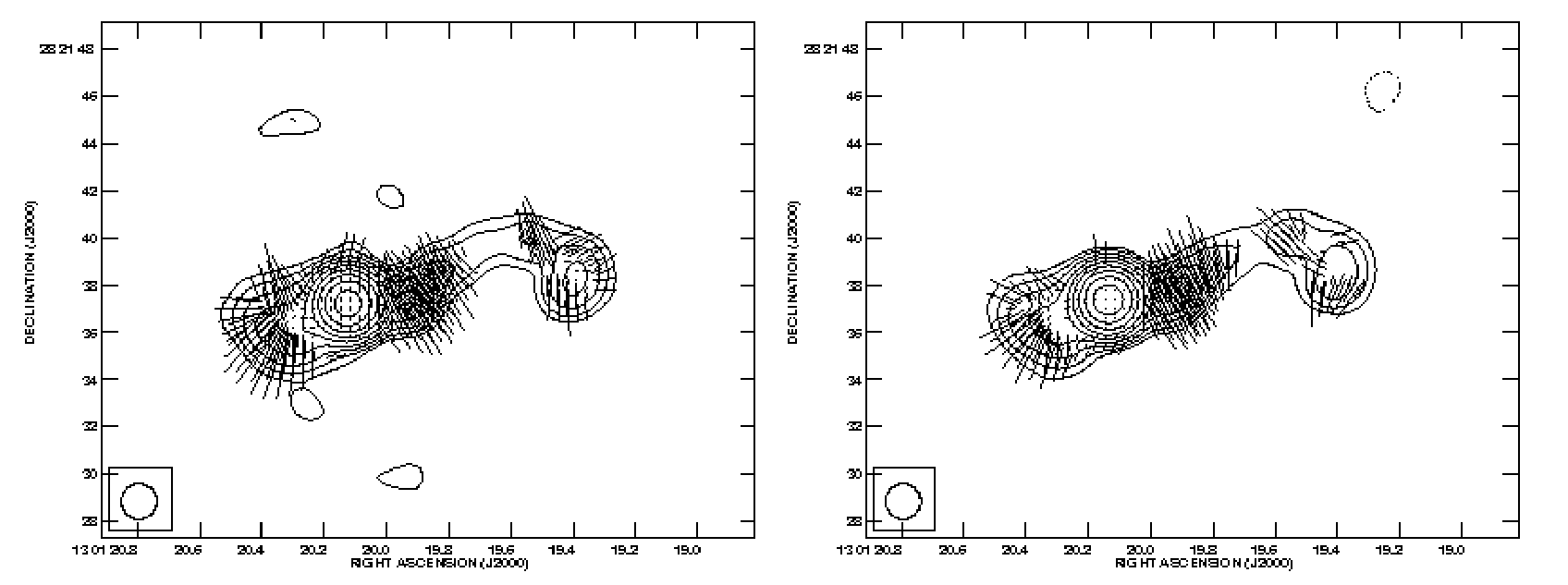}
\caption{Source 5C4.127. Total intensity radio
contours and polarization vectors at 4.535 GHz (left) and 8.465 GHz
(right). The
  bottom contour corresponds to a 3$\sigma$ noise level, contours are
  then spaced by a factor of 2. E vectors are superimposed: 
  the orientation indicates the direction of the E field, while the line
  length is proportional to the fractional polarization intensity
  (1$''$ corresponding to 10\%). }
\label{fig:127_pol}
\end{figure*}

\begin{figure*}[htb] 
\centering\includegraphics[width=0.9\textwidth]{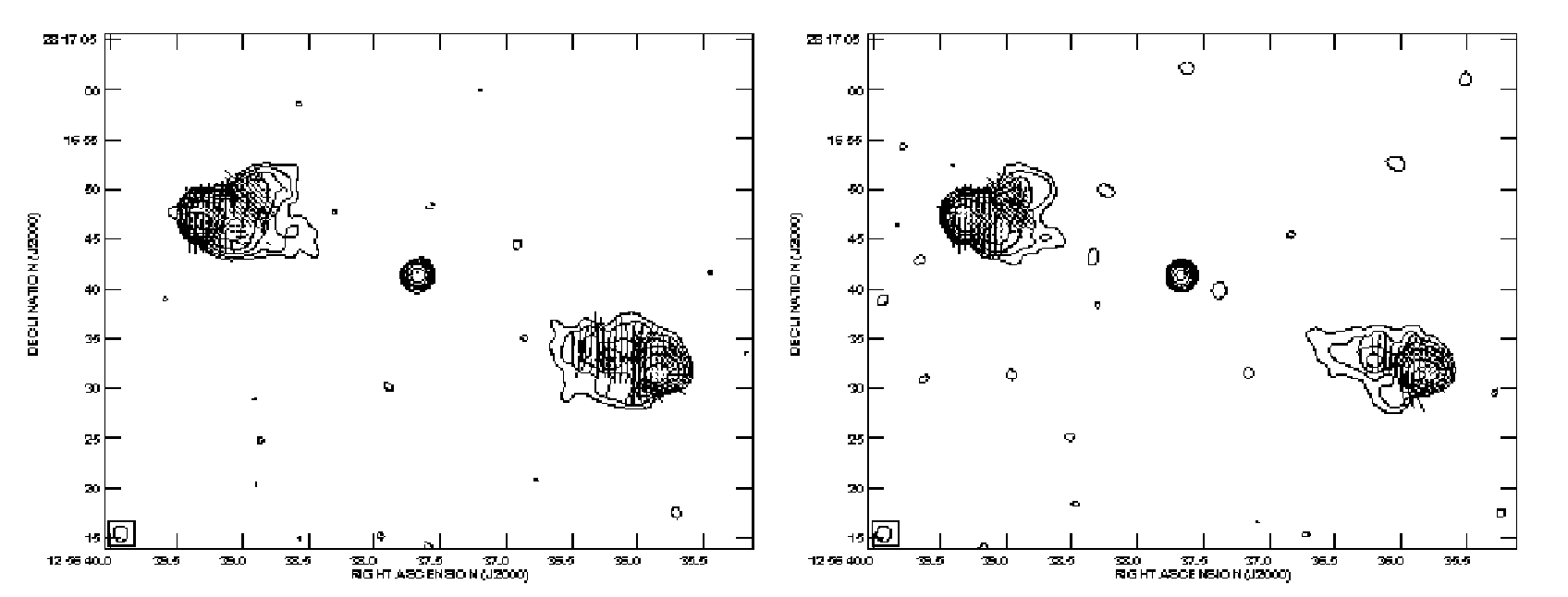}
\caption{Source 5C4.42. Total intensity radio
contours and polarization vectors at 4.535 GHz (left) and 8.465 GHz
(right). The
  bottom contour corresponds to a 3$\sigma$ noise level, contours are
  then spaced by a factor of 2. E vectors are superimposed: 
  the orientation indicates the direction of the E field, while the line
  length is proportional to the fractional polarization intensity
  (1$''$ corresponding to 10\%). }
\label{fig:42_pol}
\end{figure*}

\begin{figure*}[htb] 
\centering
\includegraphics[width=0.8\textwidth]{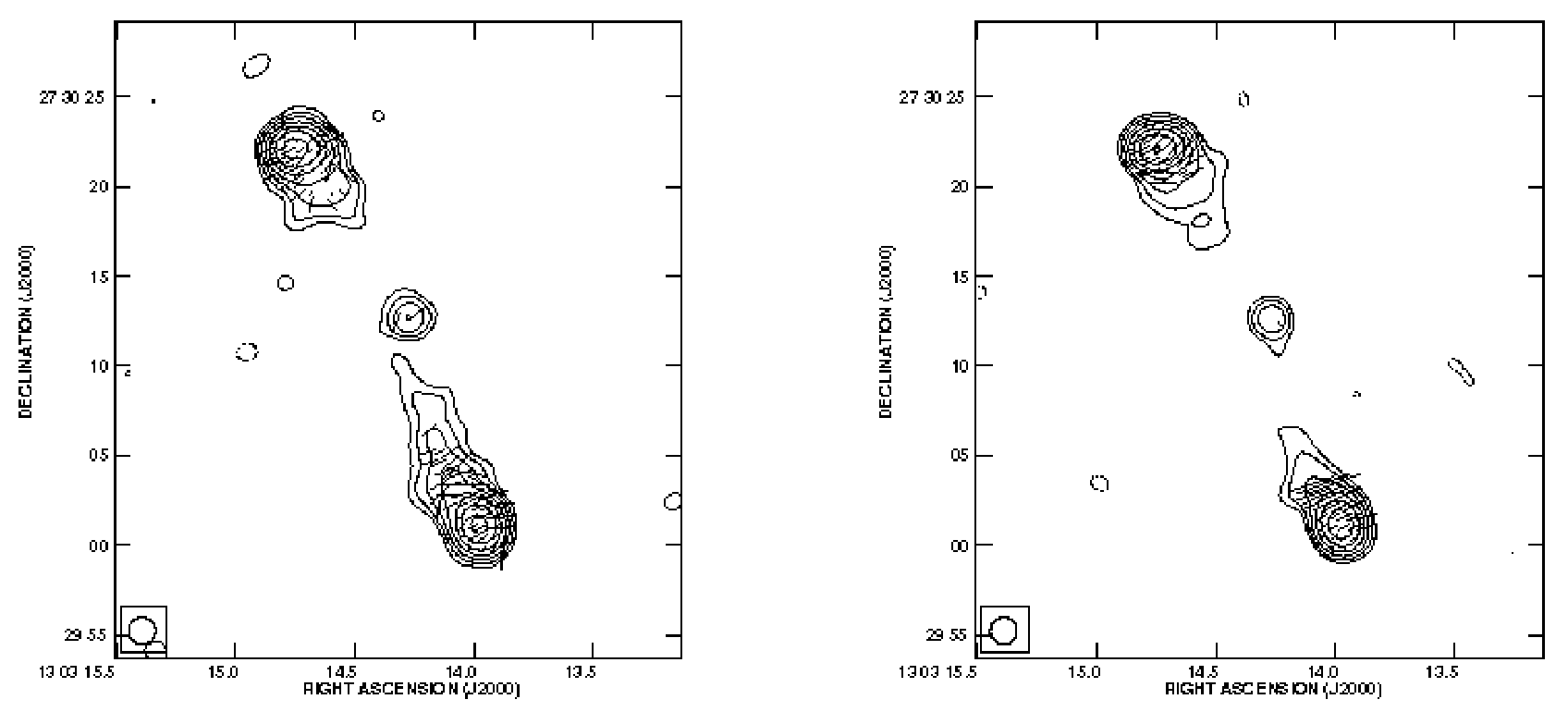}
\caption{Source 5C4.152. Total intensity radio
contours and polarization vectors at 4.535 GHz (left) and 8.465 GHz
(right). The
  bottom contour corresponds to a 3$\sigma$ noise level, contours are
  then spaced by a factor of 2. E vectors are superimposed: 
  the orientation indicates the direction of the E field, while the line
  length is proportional to the fractional polarization intensity
  (1$''$ corresponding to 10\%). }
\label{fig:152_pol}
\end{figure*}

\subsection{Radio properties of the observed sources}
\label{sec:radioSou}
In this section the radio properties of the observed sources are
briefly presented. Further details are given in Table
\ref{tab:radiomaps}.\\ Redshift information is available for three out
of the seven observed radio sources. Two of them (5C4.85 and 5C4.81)
are well studied Coma cluster members, while the third one (5C4.127)
is associated with a background source.  Although the redshift is not
known for the other four radio sources, they have not been identified
with any cluster member down to very faint optical magnitudes:
  M$_r \geq$ -15 (see Miller et al. 2009).  This indicates that
they are background radio sources, seen in projection through the
cluster. In the following the radio emission arising from the selected
sample of sources is described together with their main polarization
properties. In the fractional polarization images (from
Fig. \ref{fig:85_pol} to \ref{fig:152_pol}) pixels with errors larger
than 10\% were blanked.\\
\smallskip\\{\bf 5C4.85 - NGC 4874}\\ This a cluster source, optically
identified with the Coma central cD galaxy NGC 4874 (see {\it e.g.} Mehlert
et al. 2000).  It is a Wide Angle Tail radio galaxy, whose maximum
angular extension is $\sim$ 30$''$, corresponding to $\sim$ 15
kpc. The angular extension of the two lobes individually is larger at
the lowest frequency.  The northern lobe shows a mean fractional
polarization of 10\% and 11\% at 4.535 and 8.465 GHz respectively,
while the western lobe is less polarized ( $\sim$7\% at both
frequencies).  In Fig. \ref{fig:85_pol} the radio emission is shown
at 4.535 and 8.465 GHz.  \\
\smallskip\\ {\bf 5C4.81 - NGC 4869}\\ This source has been studied in
detail by Dallacasa et al. (1989) and Feretti et al. (1995). It is
associated with the giant elliptical galaxy NGC4869.  5C4.81 has a
Narrow Angle Tail radio morphology, and its angular size in the images
obtained here is 55$''$ (25 kpc).  The  mean fractional
polarization in the tail is $18$\% at 4.535 GHz and $21$\% at 8.465 GHz.
In Fig. \ref{fig:81_pol} the radio emission is shown at 4.535 and
8.465 GHz.\\
\smallskip\\ {\bf 5C4.74}\\ The source 5C4.74 consists of 5C4.74a and
5C4.74b, the two radio lobes of a FRII radio source. Its
redshift is unknown, and no optical identification has been found,
either with a Coma cluster member (Miller et al. 2009) nor with a
background radio source. From this we conclude that
it is a distant background source. The northeastern lobe
has a fractional polarization of $\sim$28\% and 35\% respectively at
4.535 and 8.465 GHz, while the southwestern lobe is less polarized
($\sim$ 19\% at 4.535 GHz and $\sim$ 23\% at 8.465 GHz).\\ In
Fig. \ref{fig:74_pol} the radio emission is shown at 4.535 and
8.465 GHz.\\
\smallskip\\ {\bf 5C4.114}\\ 5C4.114 is a FRI radio source, with
angular size of $\sim$ 15$''$. Its redshift is unknown, and no optical
identification either with a Coma cluster galaxy (Miller et al. 2009)
nor with a background galaxy has been found, indicating that 5C4.114
has a redshift greater than 0.023.  The southern lobe appears brighter than the
northern one. The source fractional polarization is $\sim$ 13\% at
1.365 GHz and $\sim$ 19\% at 4.935 GHz. \\ In Fig. \ref{fig:114_pol}
the radio emission is shown at 1.365 and 4.935 GHz.\\
\smallskip\\ {\bf 5C4.127}\\ 5C4.127 is a QSO located at z=1.374
(Veron-Cetty \& Veron, 2001).  Observations presented here show that
in addition to a bright nucleus the source has a weak
extension in the E-W direction of $\sim$ 16 $''$ ($\sim$ 136
kpc) at both of the observing frequency bands. The extended component
has a mean fractional polarization of $13$\% at 4.535 GHz and $14$\% at
8.465 GHz, while the nucleus is polarized at the $3$\% level.  In
Fig. \ref{fig:127_pol} radio contours of the source and polarization
vector images are shown.\\
\smallskip\\ {\bf 5C4.42}\\ 5C4.42 is a FRII-type radio
source. Redshift information is not available in the literature and no
optical identification has been found. The same arguments explained
above for the source 5C4.74 let us conclude that it is a background
radio source. The source is composed by a weakly polarized core and
two lobes that extend for $\sim$ 25$''$ in the southwest and northeast directions. The
lobes show a mean fractional polarization of $\sim$ 13\% at both 4.535
GHz and 8.465 GHz.  In Fig. \ref{fig:42_pol} radio contours and vector
polarization images of the source are shown.
\smallskip\\ 
{\bf 5C4.152}\\ 5C4.152 is a FRII type Radio Galaxy. No
redshift is available in the literature for this source. The
same arguments explained above for the source 5C4.74 let us conclude
that it is a background radio source. It is
composed of a core having a fractional polarization of a few percent
and two lobes that extend for $\sim$ 28$''$ north-south. The
lobes show a mean fractional polarization of $\sim$ $\sim$ 13\% 4.535 GHz
and  15\% at 8.275 GHz. 
In Fig. \ref{fig:152_pol} radio contours and
vector polarization images of the source are shown.

%---------------------------------------------------------------------
\section{RM: fits and errors}
\label{sec:RMobs}
\begin{figure*}[htb] 
\centering
\includegraphics[width=0.8\textwidth]{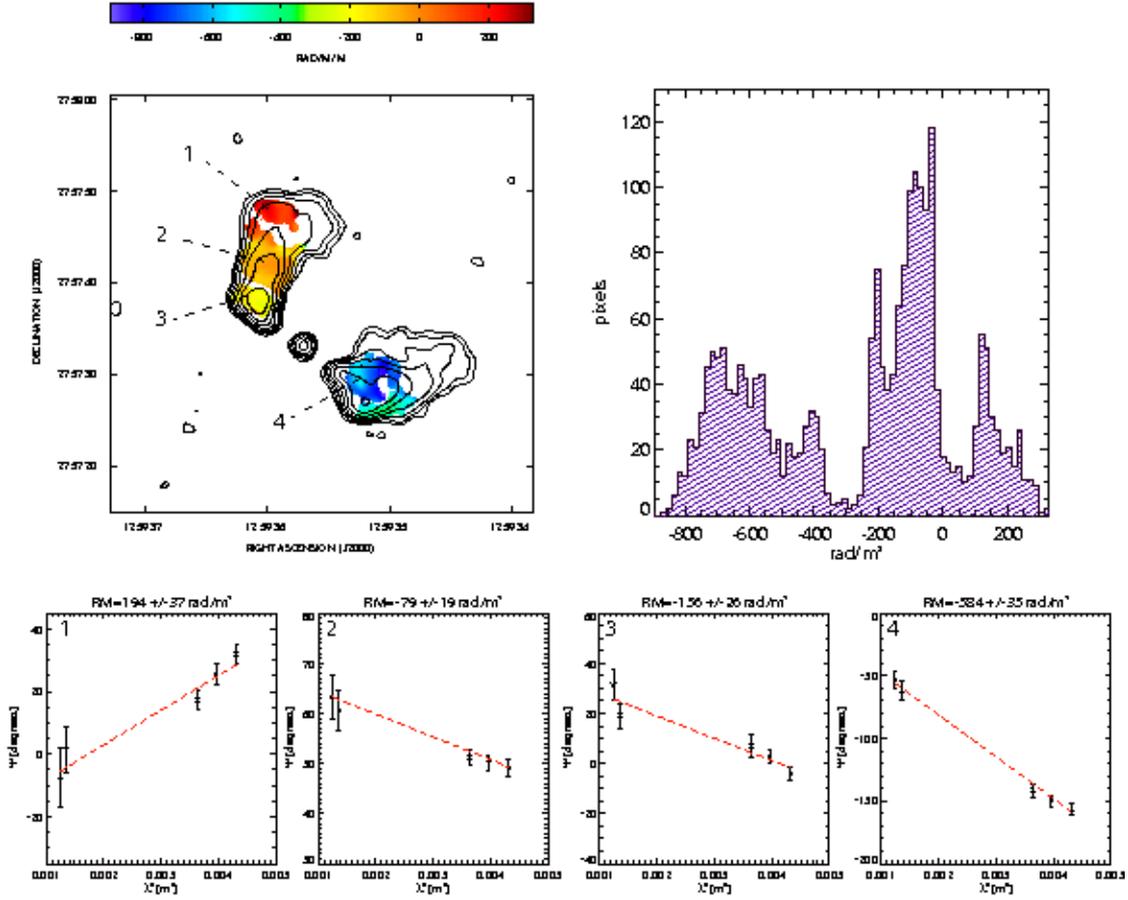}
\caption{{\bf 5C4.85:} {\it Top left:} The RM fit is shown in color along
with total intensity radio
  contours 4.935 GHz. The bottom contour correspond to the 3$\sigma$
  noise level and contours are then spaced by a factor of 2.{\it Top
    right} distribution histogram of the RM values. {\it Bottom:} fits of
  polarization angle versus $\lambda^2$ in four representative
  pixels. }
\label{fig:85_RM}
\end{figure*}

\begin{figure*}[htb] 
\centering
\includegraphics[width=0.8\textwidth]{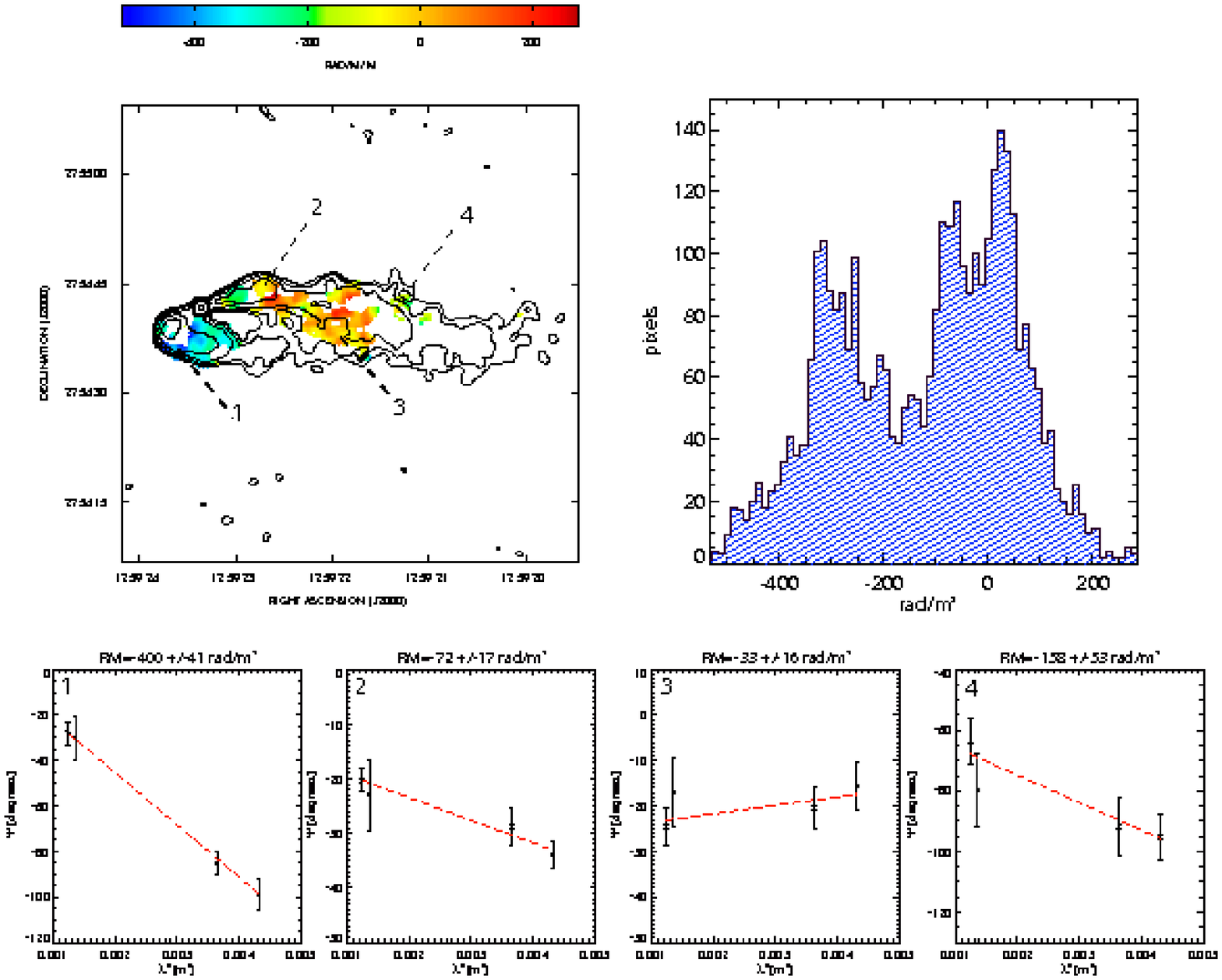}
\caption{{\bf 5C4.81:}{\it Top left:} The RM image is shown in color along with total
  intensity radio contours at 4.935 GHz. Contours start at 3$\sigma$ and 
increase by factors of 2. {\it Top right} distribution histogram of the RM
  values. {\it Bottom:} fits of polarization angle versus $\lambda^2$ in
  four representative pixels. }
\label{fig:81_RM}
\end{figure*}

\begin{figure*}[htb] 
\centering
\includegraphics[width=0.8\textwidth]{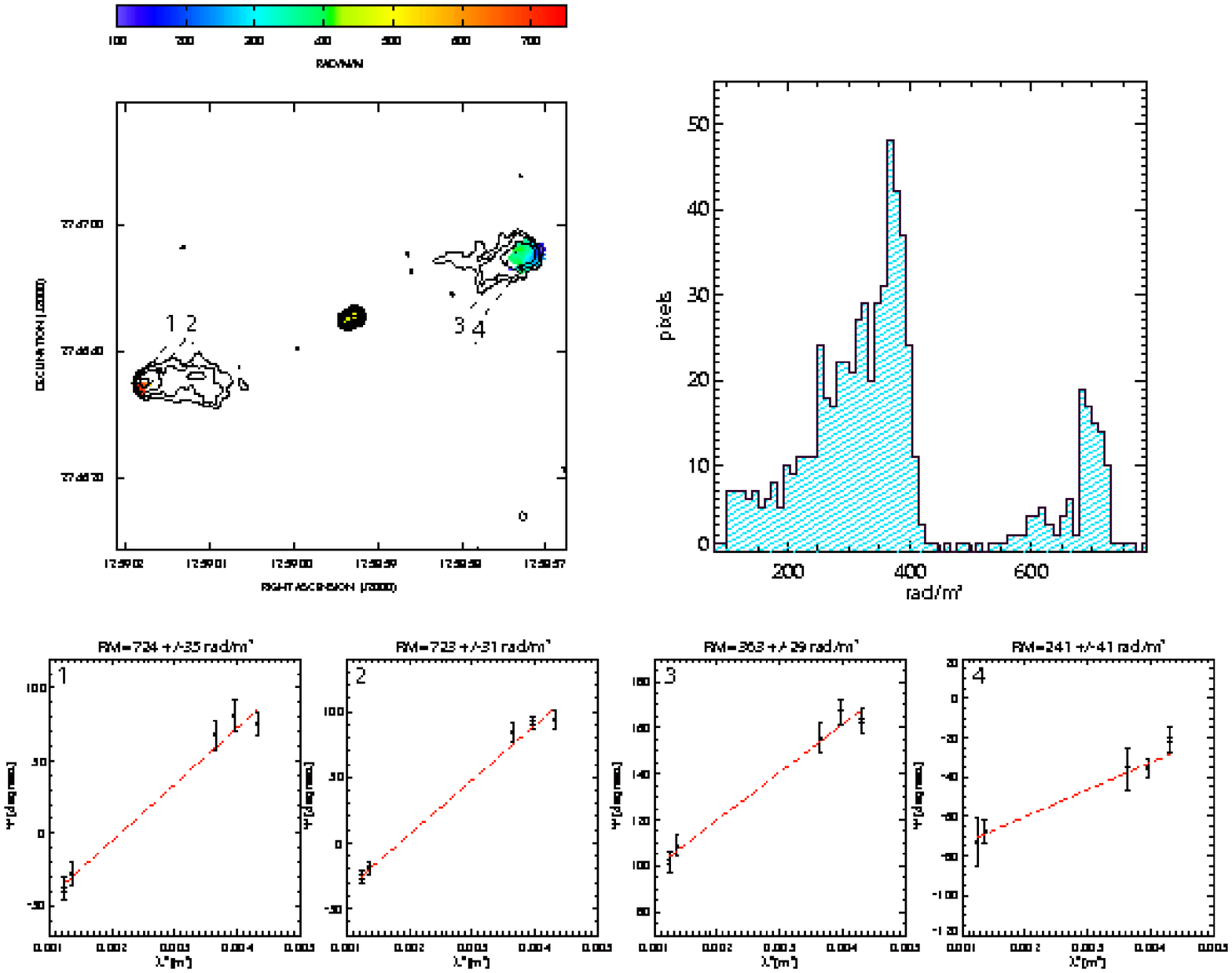}
\caption{{\bf 5C4.74:}{\it Top left:} The RM image is shown in color along
  with total intensity radio contours at 4.935 GHz. Contours start at 3$\sigma$ and 
increase by factors of 2. {\it Top right} distribution histogram of the RM values.
  {\it Bottom:} fits of polarization angle versus $\lambda^2$ in four
  representative pixels. }
\label{fig:74_RM}
\end{figure*}

\begin{figure*}[htb] 
\centering
\includegraphics[width=0.8\textwidth]{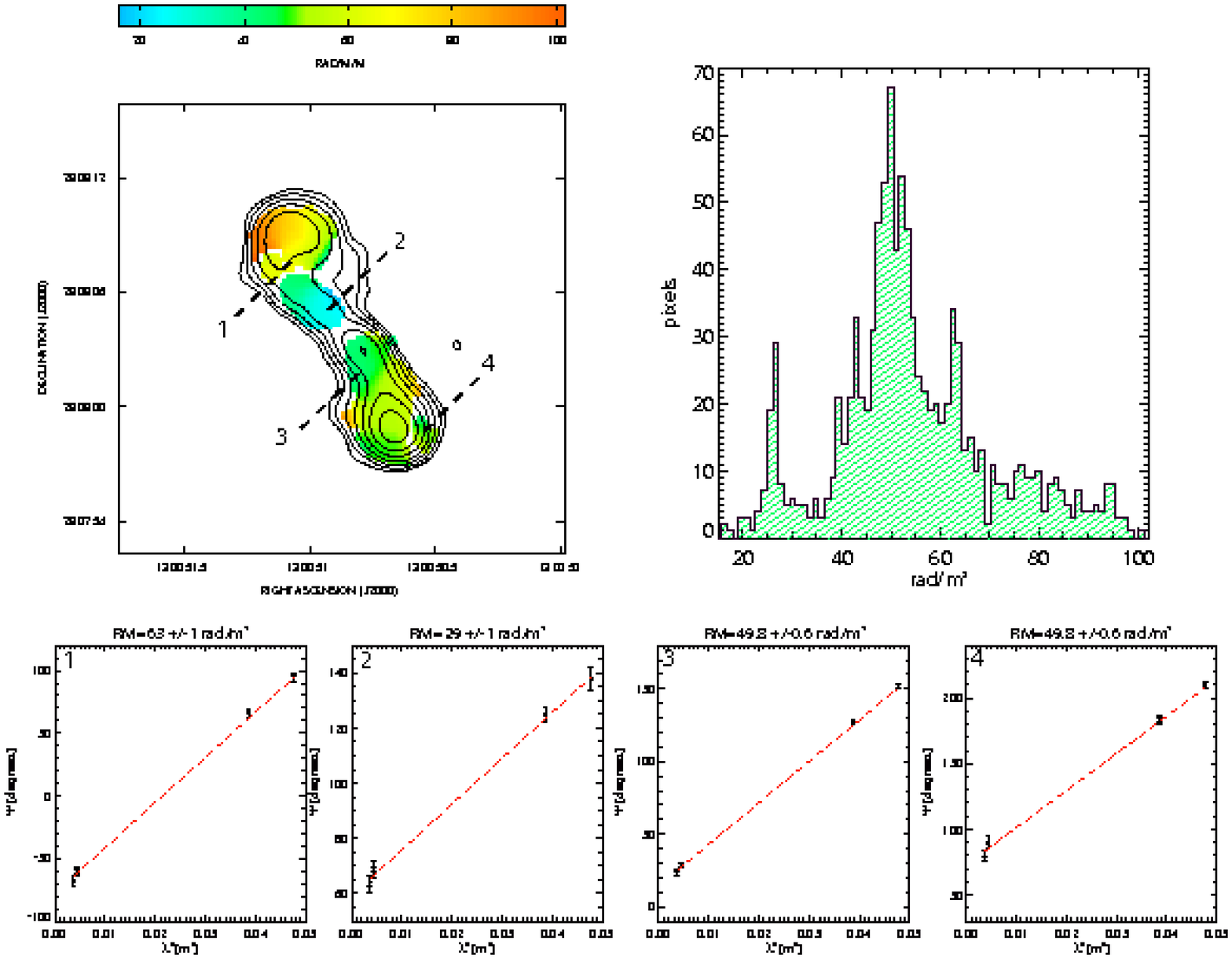}
\caption{{\bf 5C4.114:}{\it Top left:} The RM image is shown in colors along
  with total intensity radio contours at 4.935 GHz. Contours start at 3$\sigma$ and 
increase by factors of 2.  {\it Top
    right} distribution histogram of the RM values. {\it Bottom:} fits of
  polarization angle versus $\lambda^2$ in four representative
  pixels. }
\label{fig:114_RM}
\end{figure*}

\begin{figure*}[htb] 
\centering
\includegraphics[width=0.8\textwidth]{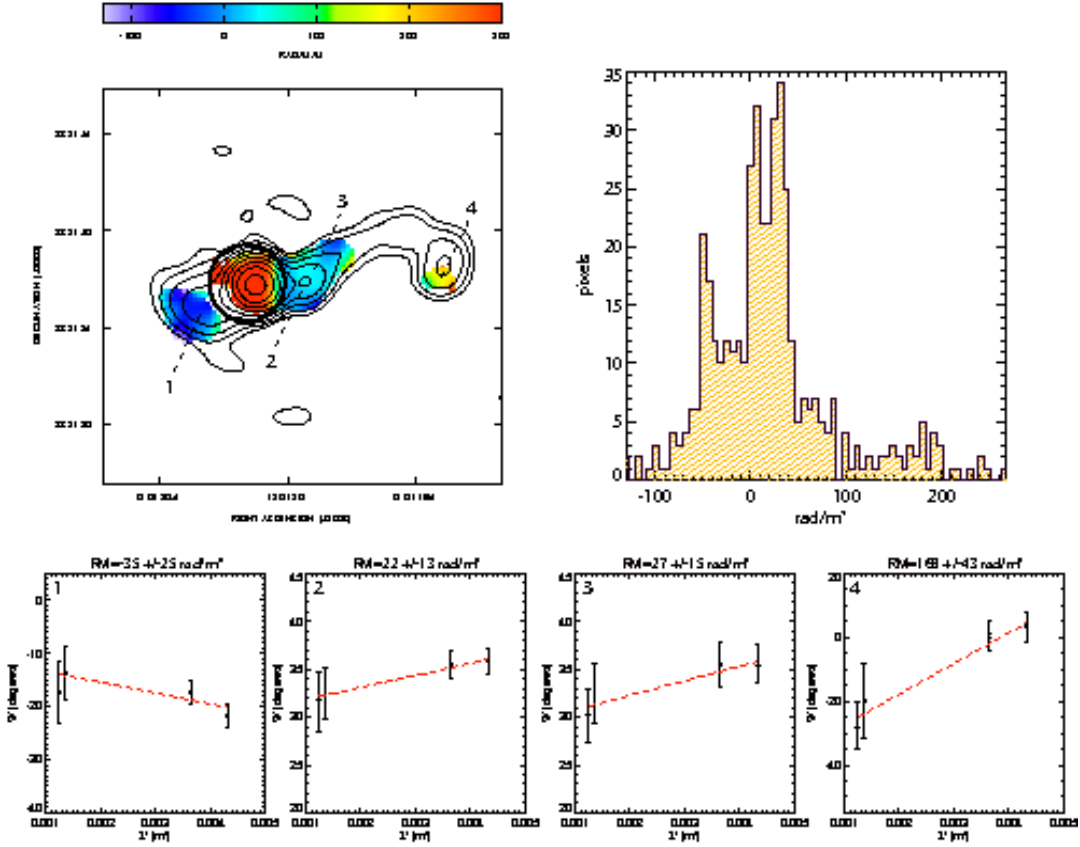}
\caption{{\bf 5C4.127:}{\it Top left:} The RM image is shown in color
  along with total intensity radio contours at 4.935 GHz. 
Contours start at 3$\sigma$ and 
increase by factors of 2.  The circle indicates the area masked in the RM
  analysis.{\it Top right} distribution histogram of the RM values.
  {\it Bottom:} fits of polarization angle versus $\lambda^2$ in four
  representative pixels. }
\label{fig:127_RM}
\end{figure*}

\begin{figure*}[htb] 
\centering
\includegraphics[width=0.8\textwidth]{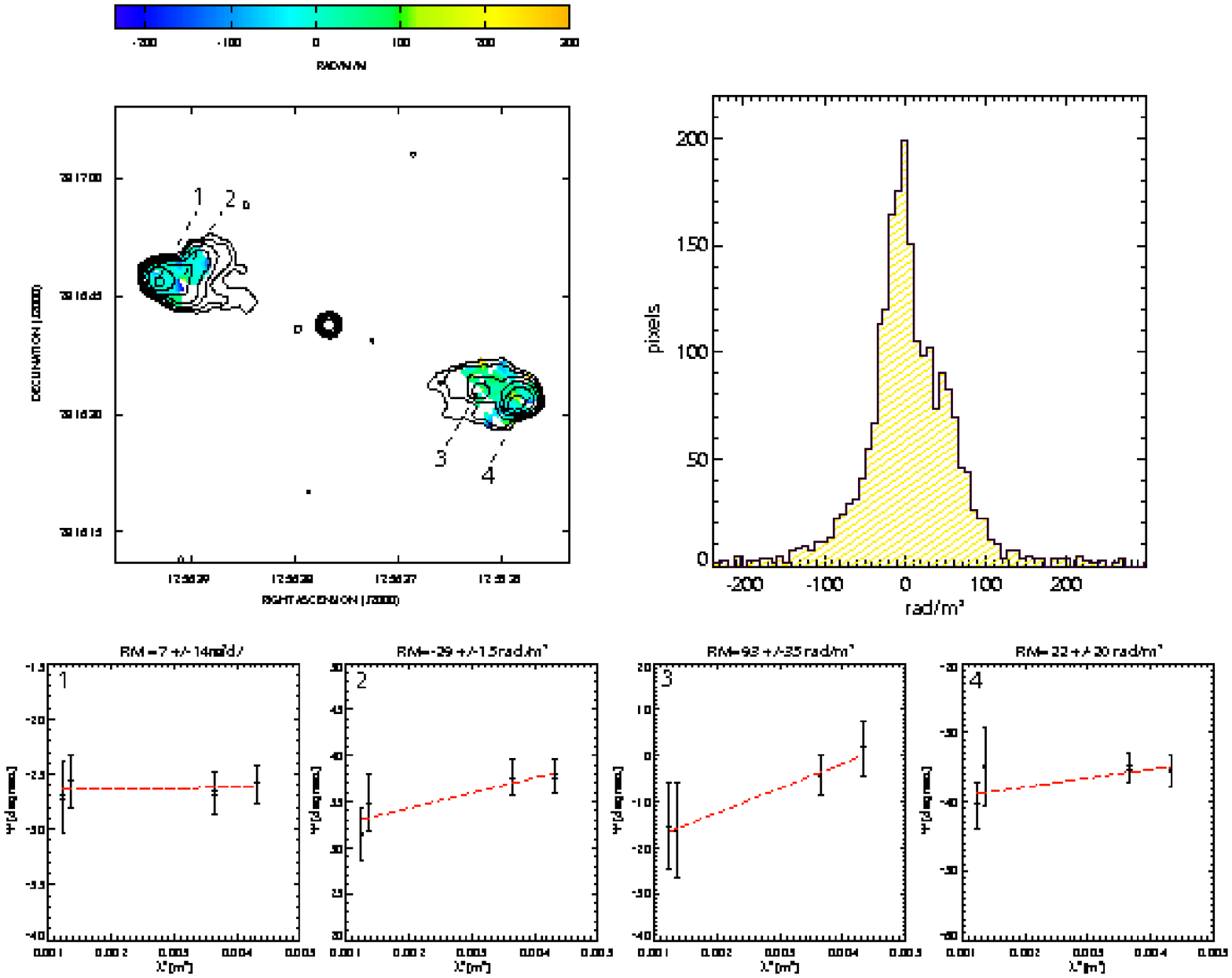}
\caption{{\bf 5C4.42:}{\it Top left:} The RM image is shown in color 
  along with total intensity radio contours at 4.935 GHz. 
Contours start at 3$\sigma$ and 
increase by factors of 2.
{\it Top right} distribution histogram of the RM values.
  {\it Bottom:} fits of polarization angle versus $\lambda^2$ in four
  representative pixels. }
\label{fig:42_RM}
\end{figure*}

\begin{figure*}[htb] 
\centering
\includegraphics[width=0.8\textwidth]{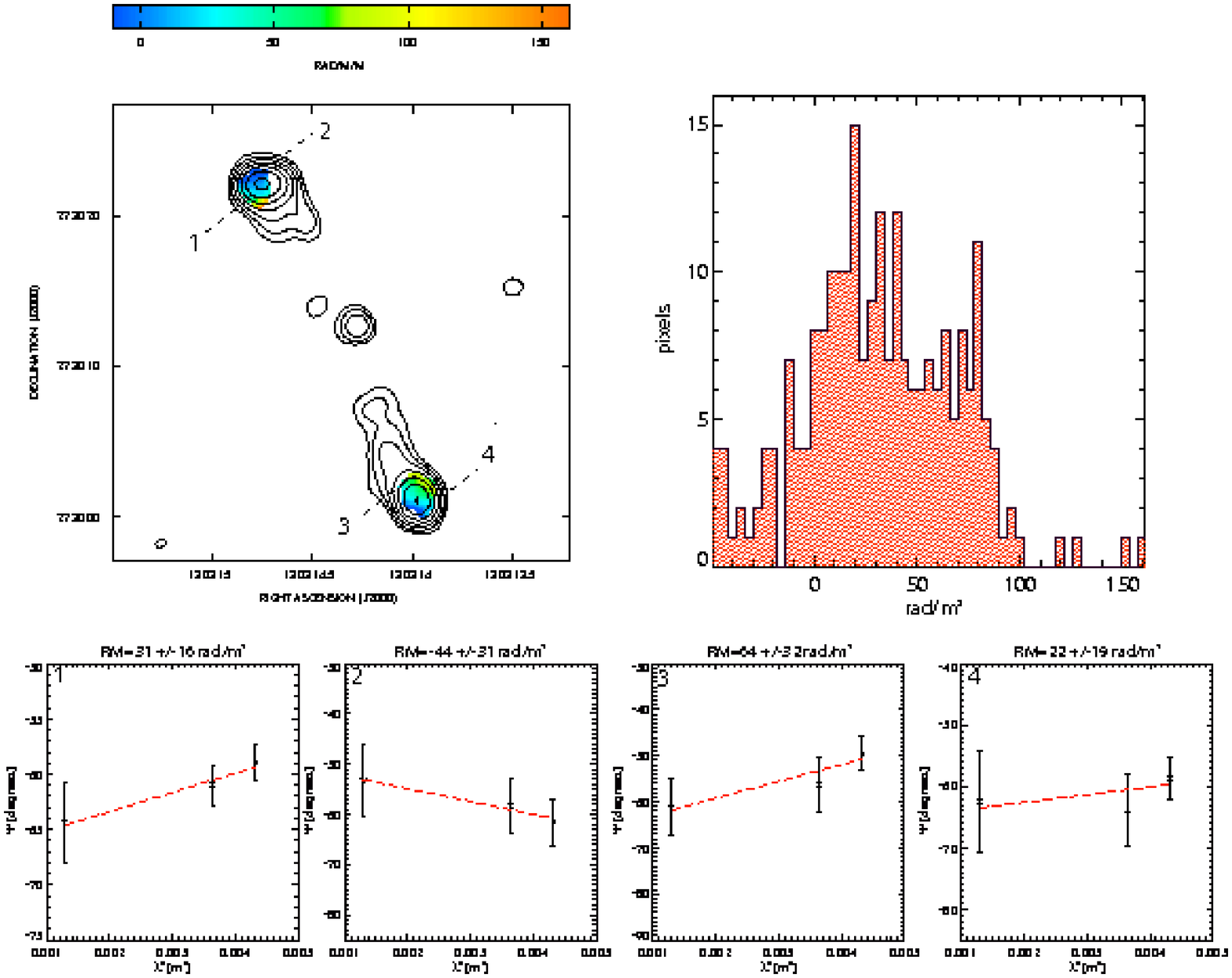}
\caption{{\bf 5C4.152:}{\it Top left:} The RM image is shown in color along
with total intensity radio contours at 4.935 GHz. 
Contours start at 3$\sigma$ and 
increase by factors of 2. {\it Top right} distribution histogram of the RM values. {\it Bottom:}
  fits of polarization angle versus $\lambda^2$ in four representative
  pixels. }
\label{fig:152_RM}
\end{figure*}

In this section the procedure used to derive the RM from the radio
observations is explained.  We used the PACERMAN algorithm
(Polarization Angle CorrEcting Rotation Measure ANalysis) developed by
Dolag et al. (2005).  The algorithm solves the n$\pi$ ambiguity in low
signal-to-noise regions by exploiting the information of nearby
reference pixels, under the assumption that the reference pixel is
connected to the nearby areas as far as the polarization angle
gradient is below a certain threshold in all of the observed frequency
maps simultaneously.\\ We considered as reference pixel those with a
polarization angle uncertainty less than 7 degrees, and fixed the
gradient threshold to 15 degrees. An error of 7 degrees in the
polarization angle corresponds to 3$\sigma$ level in both U and Q
polarization maps simultaneously. We allowed PACERMAN to perform the
RM fit if at least in 3 frequency maps the above mentioned conditions
were satisfied. The resulting RM images are shown in
Fig. \ref{fig:85_RM}, \ref{fig:81_RM}, \ref{fig:74_RM},
\ref{fig:114_RM}, \ref{fig:127_RM}, \ref{fig:42_RM} and
\ref{fig:152_RM} overlaid on the total intensity contours at 4.935
GHz. In the same figures we also provide the RM distribution
histograms and the RM fits for selected pixels in the map. The linear
trend of $\Psi$ versus $\lambda^2$ and the good fits obtained clearly
indicate that the Faraday rotation is occurring in an external
screen.\\ From the RM images we computed the RM mean ($\langle
RM\rangle$) and its dispersion ($\sigma_{RM,obs}$).  There are two
different types of errors that we have to account for: the statistical
error and the fit error. The statistical errors for $\langle
RM\rangle$ and for $\sigma_{RM,obs}$ is given by $\sigma_{RM,obs}/
\sqrt{n_b}$ and $\sigma_{RM,obs}/ \sqrt{2n_b}$ respectively, where
$n_B$ is the number of beams over which the RM has been computed. The
statistical error is the dominant one, while the error of the fit has
the effect of increasing the {\it real} $\sigma_{RM}$.  Thus, in order
to recover the {\it real} standard deviation of the observed RM
distribution we have computed the $\sigma_{RM,dec}$ as
$\sqrt{\sigma^2_{RM,obs}-Median(Err_{fit})^2}$. with
$Median(Err_{fit})$ being the median of the error distribution.
The fit error has been estimated with Monte Carlo
simulations.  We have extracted $n_B$ values, from a random Gaussian
distribution having $\sigma=\sigma_{RM,obs}$ and mean $=\langle RM
\rangle$, we have then added to the extracted values a Gaussian noise
having $\sigma_{noise}=Err_{fit}$, in order to mimic the effect of the
noise in the observed RM images. We have computed the mean and the
dispersion ($\sigma_{sim}$) of these simulated quantities and then
subtracted the noise from the dispersion obtaining
$\sigma_{sim,dec}=\sqrt{\sigma_{sim}^2-\sigma_{noise}^2}$. We have
thus obtained a distribution of $\sigma_{sim,dec}$ and means.  The
standard deviation of the $\sigma_{sim,dec}$ distribution is then the fit
error on $\sigma_{RM,dec}$ while the standard deviation of the mean
distribution is the fit error on $\langle RM\rangle$. We checked that the
mean of both distributions recover the corresponding observed values.
In Table \ref{tab:rm1} we report the RM mean, the observed RM
dispersion ($\sigma_{RM,obs}$), the value of $\sigma_{RM,dec}$
(hereafter simply $\sigma_{RM}$), with the respective errors, the
average fit error ($Err_{fit}$), and the number of beam over which the
RM statistic is computed ($n_b$).

\subsubsection{The source 5C4.74}
\label{sec:74}
The value of $\langle RM \rangle$ that we have derived for the source
5C4.74 is quite high compared with the values found for the other
sources in this cluster and it is also higher than the values obtained
in other clusters for sources at similar distances from the cluster
center ({\it e.g.} Clarke et al. 2004). The level of polarization
  of this source is also quite high compared to the other sources (see
  Sec. \ref{sec:radioSou}). We note its position southwest of the
  cluster core, in the direction of the sub-group NGC4839 that is
  currently merging with the Coma cluster (Feretti \& Neumann 2006).
  One possibility is that the magnetic field and/or the thermal gas
  has been compressed and ordered in this region, increasing the
  observed polarization flux and making $\langle RM \rangle$ peculiar
  in this position.  This might imply that more sophisticated models,
  that include deviations from a spherical symmetry could give a better
  representation of the gas density profile. We note however that the
  X-ray analysis performed in the literature by  Briel et
  al. (1992) shows that the spherical $\beta-$model is a good
  representation of the cluster X-ray surface brightness, indicating
  that deviations from spherical symmetry are small. The $\sigma_{RM}$
  value is fully compatible with the trend suggested by the other
  sources. This could be explained in the proposed scenario, if the
  magnetic field has been compressed and ordered.  In the following
  analysis we will use mainly the $\sigma_{RM}$ to infer the magnetic
  field strength and structure. Nonetheless, results will also be
  presented excluding this source from our analysis.

\subsection{Galactic contribution} 
The contribution to the Faraday RM from our Galaxy may introduce
an offset in the Faraday rotation that must be removed.  This
contribution depends on the galactic positions of the observed
sources. The Coma cluster Galactic coordinates are $l=58^{\circ}$ and
$b=88^{\circ}$. The cluster is close to the galactic north pole, so
that Galactic contribution to the observed RM is likely negligible.
However, in order to estimate this contribution the average RM for
extragalactic sources located in projection nearby the Coma cluster
region has been computed using the catalogue by Simard-Normadin et
al. (1981). The sources have been weighted by the inverse of the
distance from the Coma cluster center, and the results in a region
of 25$\times$25 degrees$^2$ centered on the cluster gives a Galactic
contribution of -0.15 rad/m$^2$. This small contribution is thus
completely negligible and has been ignored in the following
analysis. \\

\begin{table*}[htb] 
\caption{Rotation Measures values of the observed sources}          
\label{tab:rm1}      
\centering          
\begin{tabular}{|c c c c c c c|}    
\hline
\hline
Source    & Projected distance & n. of beams &$\langle RM\rangle$    &$\sigma_{RM,obs}$& Err$_{fit}$& $\sigma_{RM}$ \\
          &   kpc    &     & rad/m$^2$   & rad/m$^2$       &  rad/m$^2$ &  rad/m$^2$  \\    
\hline            
5C4.85    &   51   &  35   &-256$\pm$50  &   303    & 46     &   299$\pm$36     \\
5C4.81    &   124  &  56   &-120$\pm$22   &   166   & 48     &   159$\pm$17        \\ 
5C4.74    &   372  &  10   &372$\pm$51   &   154    & 44     &   148$\pm$41      \\ 
5C4.114   &   532  &  16   & 51$\pm$4    &    16    &  2     &    16$\pm$3     \\ 
5C4.127   &   919  &   7   & 21$\pm$30    &    65   & 36     &    54$\pm$26         \\
5C4.42    &  1250  &  33   & 6$\pm$12     &    56   & 43     &    36$\pm$11           \\ 
5C4.152   &  1489  &   4   & 32$\pm$27    &    37   & 28     &    24$\pm$21    \\
\hline
\multicolumn{7}{l}{\scriptsize Col. 1: Source name Col. 2: Source
                   projected distance from the X-ray cluster center;}\\
 \multicolumn{7}{l}{\scriptsize Col. 3:  number of beams over which RMs are computed;}\\
\multicolumn{7}{l}{\scriptsize Col. 4: Mean value of the observed RM distribution;}\\
\multicolumn{7}{l}{\scriptsize Col. 5: Dispersion of the observed RM distribution; }\\
\multicolumn{7}{l}{\scriptsize Col. 6: Median of the RM fit error; Col 7: Dispersion of the RM distribution after noise deconvolution.}
\end{tabular}
\end{table*}

\subsection{RM local contribution}
\label{sec:RMlocal}
 We discuss here the possibility that the
 RM observed in radio galaxies are not associated with the foreground
 ICM but may arise locally to the radio source (Bicknell et al. 1990,
 Rudnick \& Blundell 2003), either in a thin layer of dense warm gas
 mixed along the edge of the radio emitting plasma, or in its
 immediate surroundings. There are several arguments against
 this interpretation:
\begin{itemize}
\item{the trend of RM versus the cluster impact parameter in both
 statistical studies and individual cluster investigations (Clarke et al. 
2001, 2004; Feretti et al. 1999; Govoni et al. 2005);}\\
\item{the Laing-Garrington effect (Laing 1988, Garrington et al. 1988,
  Garrington \& Conway 1991). This effect consists of an asymmetry in
  the polarization properties of the lobes of bright radio sources
  with one-sided, large scale jets. The lobe associated with the jet
  that is beamed toward the observer is more polarized than the one
  associated with the counter-jet that points away from the
  observer. This effect can be explained if we assume that the radio
  emission from the two lobes cross different distances through the
  ICM, and therefore the emission from the counter-lobe is seen through
a greater Faraday depth, causing greater depolarization. 
This means also that the observed polarization
  properties of the source are strongly influenced by the ICM.}\\
 \item{statistical tests on the scatter plot of RM versus polarization angle
 for the radio galaxy PKS1246-410 (Ensslin et al. 2003);}\\
\item{the relation between the RM and the cooling flow rate in relaxed
 clusters (Taylor et al. 2002).}\\
\end{itemize}
This observational evidence allows us to conclude that the main
  contribution to the RM occurs in a Faraday screen located outside the
  radio sources.  The only contribution that could contaminate the
cluster Faraday screen is the contribution from the disturbed ISM in
the central parsecs of the host elliptical galaxy where the inner jet
has been found to have RMs up to thousands of radians per square meter
(Zavala \& Taylor 2004). We removed the core from the RM analysis
  in order to avoid any contribution of this kind (out to a distance
  of $\sim$5$''$ from the core). \\ The ICM origin of the observed RM
is also confirmed by the data presented here (Tab. \ref{tab:rm1}): the
trend of $\sigma_{RM}$ exhibits a decrease with increasing cluster
impact parameter. Values of $\langle RM\rangle \neq$0 and different
among sources located at different projected distances to the cluster
center indicate that the magnetic field substantially changes on
scales larger than the source size, while small RM fluctuation can be
explained by magnetic field fluctuation on scales smaller than the
source size. Thus in order to interpret correctly the RM data we have
to take into account magnetic field fluctuations over a range of
spatial scales, {\it i.e.}, we have to model the magnetic field power
spectrum.

%-------------------------------------------------------------------------
\section{The magnetic field model}
In this section the magnetic field model adopted to simulate RM images is
described. We used the approach suggested by Murgia et al. (2004),
where the magnetic field is modelled as a 3-dimensional multi-scale model and
its intensity scales with radius, following the gas distribution.
\label{sec:Sim_setup}
\subsection{The magnetic field power spectrum}
\label{sec:pssim}
In order to study the magnetic field of the Coma cluster, we
considered a 3D vectorial magnetic field model. \\ Simulations start
considering a power spectrum for the vector potential ${\bf A}$ in the
Fourier domain:
\begin{equation}
|A_k|^2 \propto k^{-\zeta}
\end{equation}
 and extract random values of its amplitude $A$ and phase $\phi$. $A$
is randomly extracted from a Rayleigh distribution (in order to
obtain a Gaussian distribution for the real magnetic field components,
as frequently observed), while $\phi$ varies randomly from 0 to 2$\pi$.
The magnetic field components in the Fourier space are then obtained
by:
\begin{equation}
{\tilde B}(k) =ik\times {\tilde A}(k).
\end{equation}
The field components $B_i$ in the real space are then derived using 3D
Fast Fourier Transform. The resulting magnetic field is a multi-scale
model with the following properties:\\ 1)$$\nabla \cdot {\bf B}=0,$$
\\ 2) the magnetic field energy density associated with each component
$B_k$ is:$$|B_k|^2 = C^2_n k^{-n},$$ $n= \zeta -2$, where $C^2_n$ is
the power spectrum normalization,\\ 3)$B_i$ has a Gaussian
distribution, with $\langle B_i\rangle=0$, $\sigma_{B_i}=\langle
B^2_i\rangle$,\\ 4) $B$ has a Maxwellian distribution.\\ The magnetic
field model scales with the radius as described in
Sec. \ref{sec:radialB}. \\ We define $\Lambda=\frac{2\pi}{k}$ as the
physical scale of the magnetic field fluctuations in the real space.
Thus in order to determine the magnetic field power spectrum in the
cluster, we have to determine three parameters: $\Lambda_{min}$,
$\Lambda_{max}$ and $n$. It is worth noting that a degeneracy arises
between $\Lambda_{max}$ and $n$ (the higher $n$ is, the lower
$\Lambda_{max}$ is, in order to produce the same RM, see also
Sec. \ref{sec:PWsim}).
\subsection{The magnetic field radial profile}
\label{sec:radialB}
There are several indications that the magnetic field intensity
decreases going from the center to the periphery of a cluster. This is
expected by magneto-hydrodynamical simulations (see e.g. Dolag et
al. 2008) and by spatial correlations found in some clusters between
thermal and non-thermal energy densities (Govoni et al. 2001). \\ We
assume that the cluster magnetic field follows the thermal component
radial distribution according to:
\begin{equation}
\langle {\bf B}\rangle(r)= \langle {\bf B}_0\rangle \left( \frac{n_e(r)}{n_0} \right)^{\eta}
\label{eq:BProfile}
\end{equation}
where $\langle {\bf B}_0\rangle$ is the mean magnetic field strength at the
cluster center. \\ In order to obtain the desired magnetic field
radial profile we have operated directly in the real space. Strictly,
this operation should be performed in the Fourier space, by convolving
the spectral potential components with the shaping profile. It has
been proved that these two approaches give negligible differences
(see Murgia et al. 2004).\\ When the magnetic field profile is
considered, two more parameters have to be determined: $\eta$ and
  $\langle {\bf B}_0\rangle$. These two parameters
are degenerate (the higher is $\langle {\bf B}_0\rangle$ the higher is $\eta$). In
fact higher central value of the field require a steeper spectrum in
order to produce the same RM.\\
\smallskip

The adopted magnetic field model has then {\it a total of 5 free
  parameters: $\Lambda_{min}$, $\Lambda_{max}$, n, $\eta$ and $\langle
  B_0\rangle$, and is subject to two degeneracies: $\Lambda_{max}$- n
  and $\eta$ and $\langle B_0\rangle$}. \\ Fitting all of these five
parameters simultaneously would be the best way to proceed, but it is
not feasible here, due to the computational burden caused by the
Fourier Transform inversion. Indeed we have to simulate a large volume
$\sim 3^3$ Mpc$^3$ with a sub-kiloparsec pixel-size. \\ The aim of
this work is to constrain the magnetic field radial profile, and for
this reason the sources were selected in order to sample different
regions at different impact parameters. This allows us to reach a good
sensitivity to the RM at different distances from the cluster
center.\\ We proceed as follows: we perform 2D simulations with
different magnetic field power spectra in order to recover the RM
statistical indicator that are sensitive to the magnetic field power
spectrum (Sec. \ref{sec:PWsim}). From this analysis we derive the
power spectrum that best reproduces the observations. We then perform
3D magnetic field simulations varying the values of $B_{0}$ and $\eta$
and derive the magnetic field profile that best reproduces the RM
observations (Sec. \ref{sec:Bprof_sim}).
\begin{figure}[h] 
\centering
\includegraphics[width=8cm]{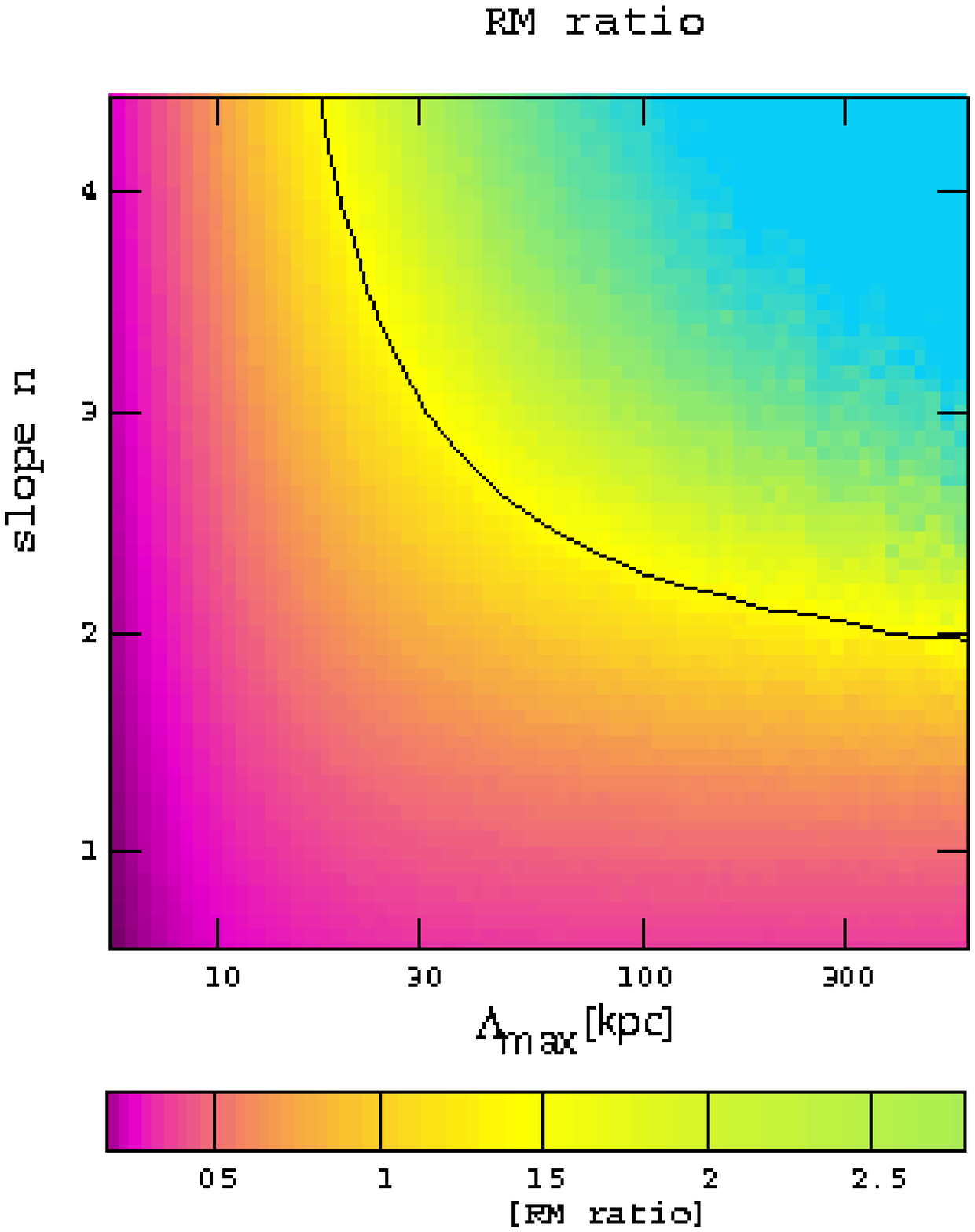}
\caption{The RM ratio $|\langle RM \rangle| / \sigma_{RM}$ as a function
of $n$ and $\Lambda_{max}$ computed on simulated RM images. The line
refers to the mean values  obtained by averaging the ratio of the
sources.}
\label{fig:lambdan}
\end{figure}
%-----------------------------------------------------------------------
\section{Comparing observations and simulations}
\label{sec:errors}
A tricky point when observations and simulations are compared is the
correct evaluations of the errors and uncertainties that this process
is subject to.  The simulations we present in this paper start from a
random seed and generate 2D and 3D magnetic fields. From these
fields simulated RM images are obtained, and then compared with those
observed in order to constrain the magnetic field properties.  
It is worth noting that for a given magnetic field
model, the RM in a given position of the cluster varies
depending on the initial seed of the simulation, so that different
realizations of the same model will correspond to different values of
$\langle RM \rangle$ and $\sigma_{RM}$ at that position. Given the
limited area covered by RM observations, the random nature of the
field cannot be neglected in our analysis. \\ We adopt the following
approach to compare observations and simulations: once the simulated
RM image is obtained for a source, it is convolved with a Gaussian
function having $FWHM$ equal to the beam $FWHM$ of the observed
image. The simulated RM image is then blanked in the same way as the observed
RM image. This ensures that simulations are subject to the same sampling
bias that we have to deal with when obtaining the RM from
observations. The comparison between the observed RM images and those
simulated is performed with the $\chi^2$ distribution, by
computing:
\begin{equation}
\label{eq:chi2}
\chi^2 = \sum^7_{i=1}{\frac{(C_{i,obs} - \langle C_{i,sim} \rangle)^2}{Err^2_{C_{i,obs}}}}
\end{equation}
where i indicates the source, $C_{obs}$ refers to a generic observed
quantity, while $\langle C_{i,sim}\rangle $ is the same quantity
averaged over the different equivalent numerical simulation that start
with different random seeds, and $Err^2_{C_{i,obs}}$ refers to the
error of $C_{obs}$. \\

%----------------------------------------------------------------------

\begin{figure*}[htb] 
\centering
\includegraphics[width=\textwidth]{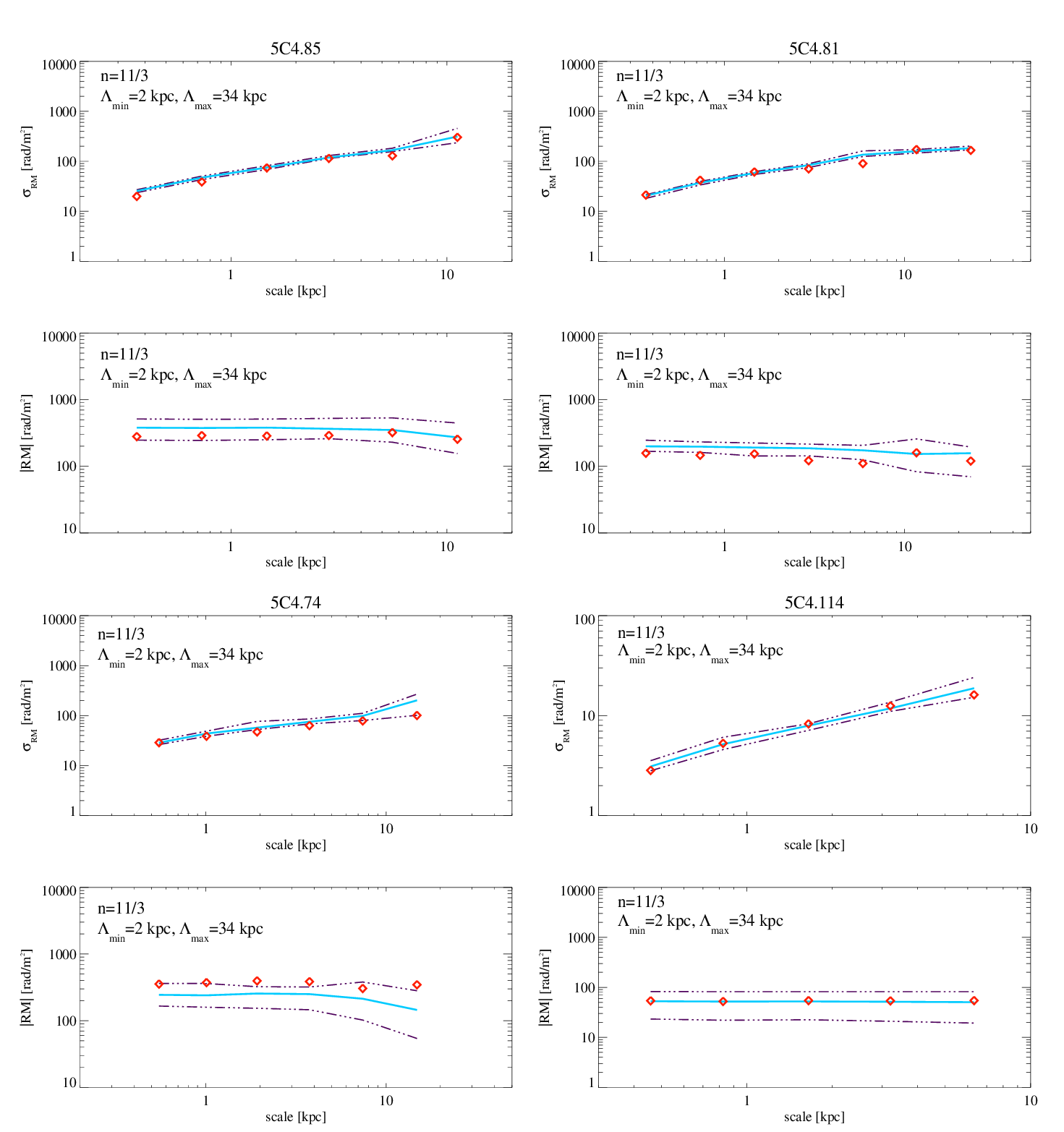}
\caption{Fits to the RM images for the Kolmogorov power spectrum that
  best reproduces the observed RM (n=11/3, $\Lambda_{min}=$2 kpc,
  $\Lambda_{max}=$34 kpc) globally  for the sources used in the 2D analysis
  (see Sec. \ref{sec:PWsim}). From top to bottom: fit to the 
  $\sigma_{RM}$ and $\langle RM \rangle$. Red diamonds represent the
  observed statistic, the cyan line represents the mean taken over ten
  different realizations of the same magnetic field power spectrum,
  and the blue lines represent the scatter in the simulations.}
\label{fig:PS_MSS}
\end{figure*}

\begin{figure*}[htb] 
\centering
\includegraphics[width=\textwidth]{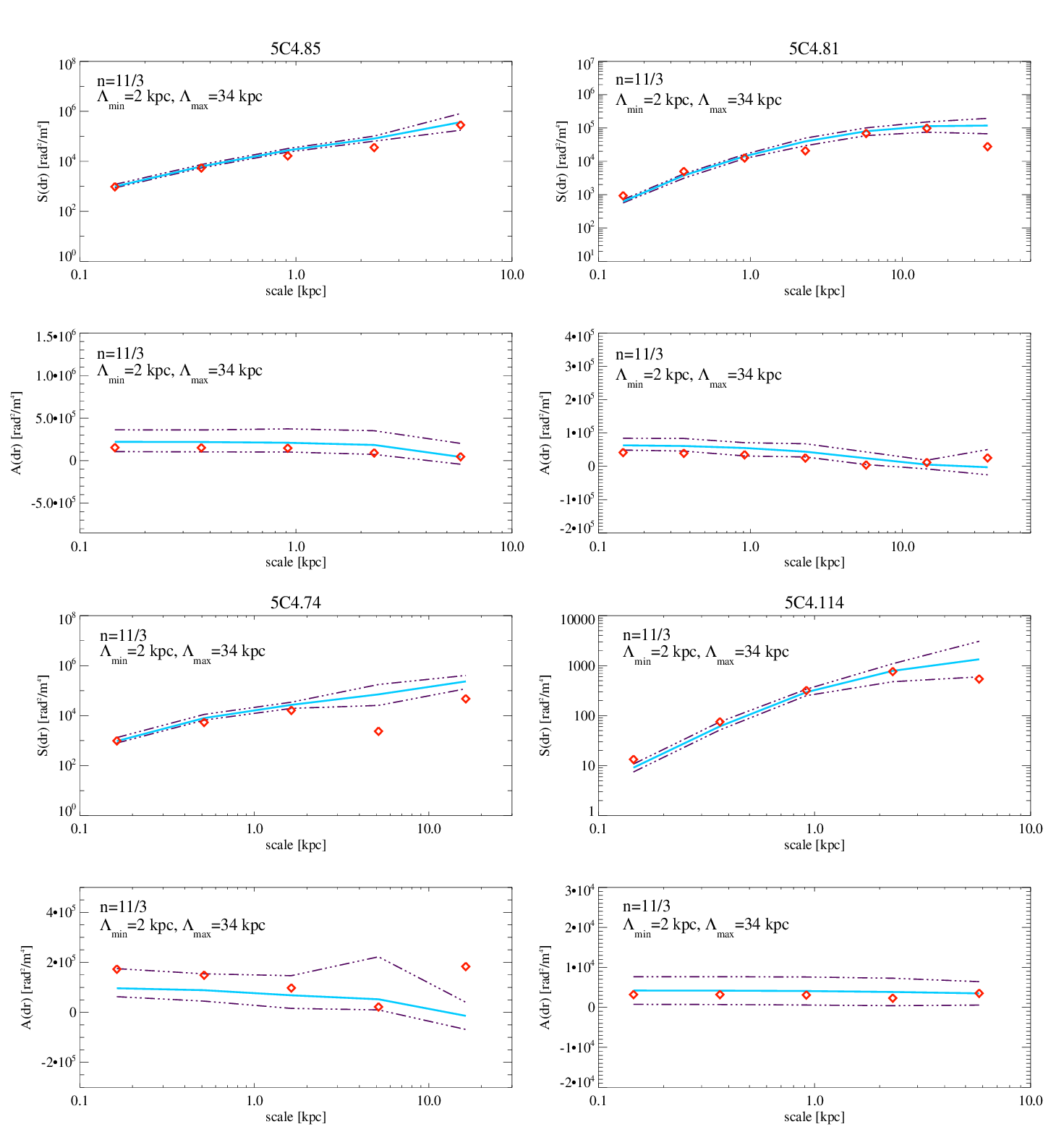}
\caption{Fits to the RM images for the Kolmogorov power spectrum that
  best reproduces the observed RM ($n$=11/3, $\Lambda_{min}=$2 kpc,
  $\Lambda_{max}=$34 kpc) globally for the sources used in the 2D analysis
  (see Sec. \ref{sec:PWsim}). Fit to the $S(r)$ (top) and $A(r)$ (bottom). Red
diamonds represent the observed statistic, the cyan line represents the
mean taken over ten different realizations of the same magnetic field
power spectrum, and the blue lines represent the scatter in the simulations.}
\label{fig:PS_SFUNC}
\end{figure*}

\section{Determining the magnetic field from RM observations}
Here we describe how the magnetic field power spectrum has
been investigated.
\label{sec:Sim}
\subsection{{\bf Constraining the magnetic field power spectrum}}
\label{sec:PWsim}
Several observational quantities can be useful to constrain some
properties of the magnetic field power spectrum. In particular:
\begin{itemize}
\item{Both $\langle RM \rangle$ and $\sigma_{RM}$ scale linearly with
the magnetic field strength, while they have different trends with $n$ and
$\Lambda_{max}$, which are degenerate parameters. The ratio $|\langle
RM \rangle|/\sigma_{RM}$ can thus be used to investigate the magnetic
field power spectrum (see also Fig. 3 in Murgia et al. 2004).}
\item{The minimum scale of the magnetic field fluctuation,
  $\Lambda_{min}$, affects the depolarization ratio (DP ratio) at two
  different frequencies
  (i.e. $DP=\frac{P_{\nu_1}/I_{\nu_1}}{P_{\nu_2}/I_{\nu_2}}$) and the
  $\sigma_{RM}$. Both $DP$ and $\sigma_{RM}$ are in fact determined by
  the magnetic power on the small spatial scales. This parameter can
  be thus be derived by studying high resolution polarization images.}
\item{It has been demonstrated that the magnetic field auto-correlation
function is proportional to the RM auto-correlation function 
(Ensslin \& Vogt 2003). Since the power spectrum is the
Fourier transform of the auto-correlation function, it is possible to
study the 3D magnetic field power spectrum starting from the power
spectrum of the RM images.}

\end{itemize}
We simulated 2D magnetic field models with different power spectra and
compared simulated RM images and DP with the corresponding observable
quantities. In these simulations the power spectrum normalization is
set independently for each source\footnote{In these 2D RM simulations
  the radial profile of the magnetic field is not accounted for. This
  implicitly assumes that the mean magnetic field strength is not
  dramatically varying over the scale of the source. This is a
  reasonable assumption since the linear sizes of the sources at the
  Coma redshift are in fact much smaller that the cluster core
  radius.}.  The computational grid is $512 \times 512$ pixel$^2$ and
the pixel-size was fixed to 0.2 kpc. This guarantees that each beam is
represented by three pixels in the grid.  The resulting field of view
is then $\sim$ 100$\times$100 kpc$^2$, that is enough to recover the
projected size of the sources and to properly sample the large power
spectrum scales.  Simulations were convolved with a Gaussian function
having FMHM equal to the beam of the observations.

\subsubsection{\bf The $\Lambda_{max}$-n plane}
\label{sec:lambdan}
In order to illustrate the degeneracy existing between $\Lambda_{max}$
and $n$, 2D magnetic field models with different power
spectra have been simulated. We allowed the parameter $n$ to vary in
the range $[0.5;4.5]$ and $\Lambda_{max}$ in the range $[5;600]$ kpc,
with steps of 0.06 and 9.3 kpc respectively. We derived simulated RM
images for each combination of these parameters and calculated the RM
ratio as:
 \begin{equation}
 RM ratio=\frac{|\langle RM \rangle| } {\sigma_{RM} }
 \end{equation}
in a region of $15\times15$ kpc$^2$, comparable to the regions
where RM has been observed.  In Fig. \ref{fig:lambdan} the values of
the RM ratio for the simulated RM images is shown in colors, as a
function of both $n$ and $\Lambda_{max}$.  The RM ratio was computed for
the observed source that have a RM signal-to noise ratio $>$3 both for
$\sigma_{RM}$ and $\langle RM \rangle$, i.e for the sources 5C4.85,
5C4.81, 5C4.74 and 5C4.114.   The
resulting values are shown in Fig. \ref{fig:lambdan} (black
line).\\ The plot shown in Fig.\ref{fig:lambdan} shows what
$\Lambda_{max}-n$ degeneracy means: the same value of the RM ratio
can be explained with different power spectra.  There are, as expected,
two asymptotic trends. In fact, if the magnetic field power spectrum
is flat (e. g. $n\leq$2), the bulk of the magnetic field energy is on the
small scales, and thus the effect of increasing $\Lambda_{max}$ is
negligible after a certain threshold, that in this case is achieved for
$\Lambda_{max}\sim $300 kpc.  As the power spectrum steepens the bulk
of the energy moves to large scales, and thus as $\Lambda_{max}$
increases, the energy content also increases sharply. This is the
reason of the second asymptotic trend that is shown in the plot: as $n$
increases $\Lambda_{max}$ decreases faster and faster. As $n$ approaches
the value of $\sim$ 11/3 (Kolmogorov power spectrum), the observed
data constrain $\Lambda_{max}$ to be $\sim$20 - 40 kpc.\\

\begin{figure}[h]
\centering
\includegraphics[width=9cm]{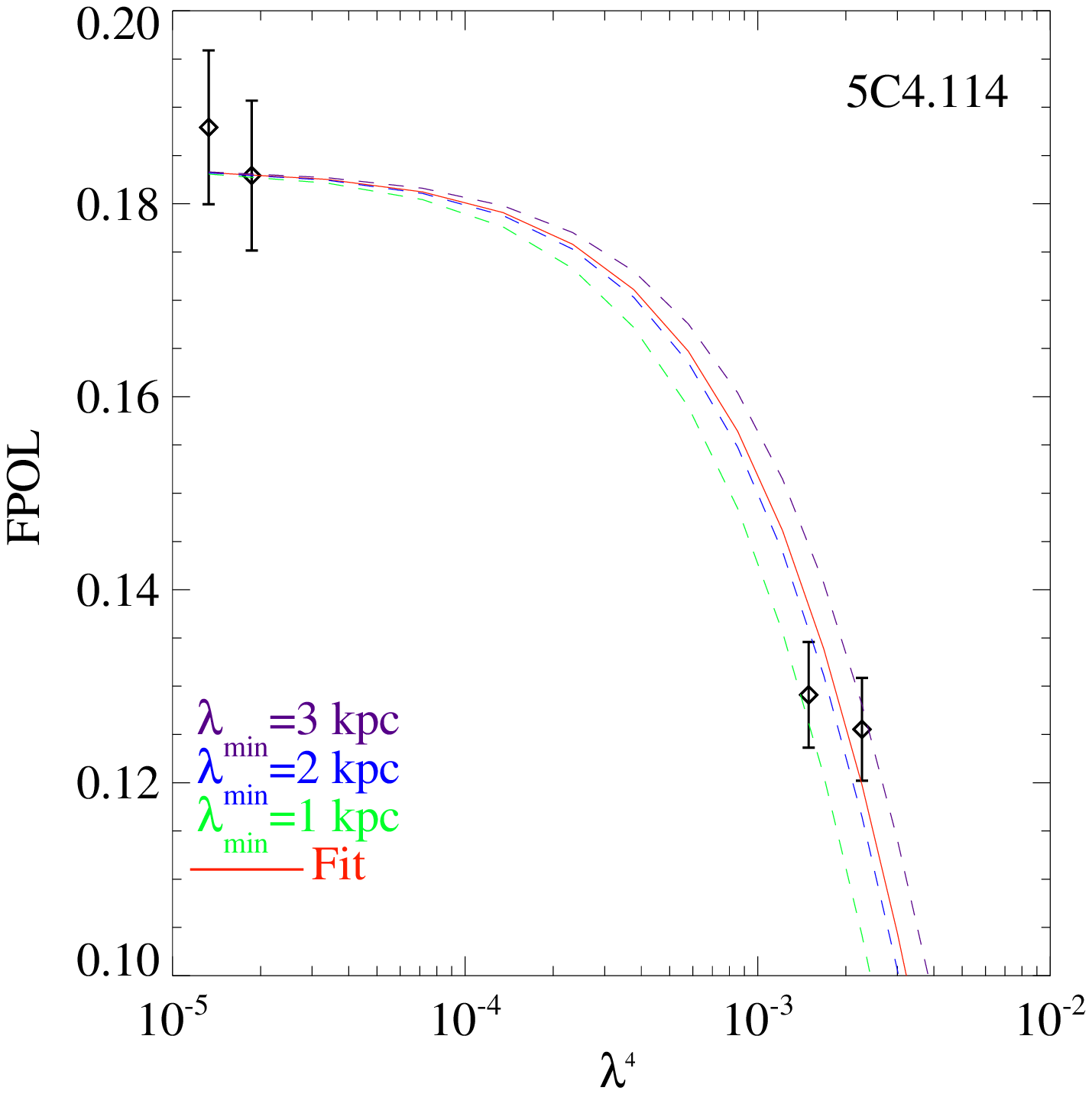}
\caption{Fits to the Burn law. Points refer to observed data, while the red
  line is the fit obtained from observations. Dashed lines refer to
  the fits obtained from three different models, with different values
  of $\Lambda_{min}$, as reported in the bottom left corner of the
  plot.}
\label{fig:114kburn}
\end{figure}

\subsection{Structure function, auto-correlation function and Multi-Scale-Statistic}
In order to constrain more precisely the estimate of the magnetic
field power spectrum parameters indicated by the previous analysis we
have investigated the statistical properties of the RM images
individually.  We have fixed $n=11/3$, corresponding to the Kolmogorov
power law for turbulent fields. This choice is motivated by both
observational and theoretical works. Schuecker et al. (2004) analyzed
spatially-resolved gas pseudo-pressure maps of the Coma galaxy cluster
deriving that pressure fluctuations in the cluster center are
consistent with a Kolmogorov-like power spectrum. Furthermore,
cosmological numerical simulations have recently demonstrated that 3D
power spectrum of the velocity field is well described by a single
power law out to at least one virial radius, with a slope very close to
the Kolmogorov power law (Vazza et al. 2009a, b). \\ The range of
values of $\Lambda_{max}$ is suggested by the previous analysis (see
Fig. \ref{fig:lambdan}). In order to choose the best parameters in
that range, and to find the best value for $\Lambda_{min}$, we
simulated RM images and used two different statistical methods to
compare the observed RM images to the simulated ones:
\begin{enumerate}
\item{ We calculated the auto-correlation function and the structure
  function of the observed RM image, and then compared them with the
  simulated RM image.  The RM structure function is defined as
  follows.
\begin{equation}
S(dx,dy)=\langle [RM(x,y)-RM(x+dx,y+dy)]^2\rangle_{(x,y)},
\end{equation}
  where $=\langle \rangle_{(x,y)}$ indicates that the average is taken
  over all the positions $(x,y)$ in the RM image. Blank pixels were
  not considered in the statistics. The structure function $S(r)$ is
  then computed by radially averaging $S(dx,dy)$ over regions of
  increasing size of radius $r=\sqrt{dx^2+dy^2}$. $S(r)$ is thus
  sensitive to the observable quantity $\sigma_{RM}$ over different
  scales.  The auto-correlation function is defined as:
\begin{equation}
A(dx,dy)=\langle [RM(x,y) RM(x+dx,y+dy)]\rangle_{(x,y)}
\end{equation}
Since $A(0)=\langle  RM^2 \rangle=\sigma^2_{RM}+\langle  RM \rangle^2$,
the auto-correlation function is sensitive to both $\langle  RM
\rangle$ and the $\sigma_{RM}$.}
\item{We computed a Multi-Scale Statistic, namely we computed $\langle
  RM \rangle$ and $\sigma_{RM}$ over regions of increasing size in the
  observed RM images and compared them with the same values obtained
  in the simulated images. The smallest region over which $\langle RM
  \rangle$ and $\sigma_{RM}$ are computed corresponds to a box of 0.4
  $\times$ 0.4 kpc size. The box side is then increased by a factor
  two until the full source size is reached.  We note that this
  approach is sensitive to both $\langle RM \rangle$ and $\sigma_{RM}$
  over different spatial scales, and is thus a useful tool to
  discriminate among different power spectra. This indicator differs
  from the $S(r)$ and $A(r)$ in that as $r$ increases, the number of
  pixels useful for computing the Multi-Scale Statistic increases,
  giving a robust statistical estimate on large scales. }
\end{enumerate}
 For each source we simulated different power spectra varying
  $\Lambda_{min}$ from 1 kpc to 5  kpc and $\Lambda_{max}$ from
  20 to 40 kpc.  For every power spectrum and for each source we
  realized ten different 2D RM images, and compared the statistics
  ($S(r)$, $A(r)$ and Multi-Scale Statistic) with the observed ones by
  computing Eq. \ref{eq:chi2}.
This approach makes it possible to 
discriminate the best power spectrum model 
compatible with our data. In this 2D analysis we focused on the
sources 5C4.85, 5C4.81, 5C4.74 and 5C4.114, whose RM images have
signal-to-noise ratio $>$3. Each source was fitted separately.  
The total $\chi^2$ was then computed by summing the individual values
obtained for each source.  The minimum value of the total
$\chi^2_{tot}$ corresponds to the power spectrum model characterized
by $\Lambda_{max}=34$ kpc and $\Lambda_{min}=2$ kpc.  We show in
Figs. \ref{fig:PS_MSS} and \ref{fig:PS_SFUNC} the Kolmogorov power
spectrum model that best reproduces the observed RM images. In
Appendix \ref{sec:app} similar Figures obtained with other power
spectrum models are shown.

\subsubsection{\bf $\Lambda_{min}$ and Fractional polarization}
It has been demonstrated (Burn 1966, see also Laing 2008) that when
Faraday Rotation occurs the fractional polarization $FPOL$ can be related
to the fourth power of the observing wavelength $\lambda$
according to the Burn law:
\begin{equation}
\label{eq:kburn}
FPOL=\frac{P_{\lambda}}{I_{\lambda}}\propto exp(-k \lambda^4). 
\end{equation}
Since$FPOL$ is sensitive to the minimum scale of the power spectrum,
$\Lambda_{min}$, Eq. \ref{eq:kburn} can be used to constrain it.  We
fitted Eq. \ref{eq:kburn} to our observations and to our simulations,
performed with different values of $\Lambda_{min}$. These fits
indicate that a Kolmogorov power spectrum with the best agreement with
observations is achieved for $\Lambda_{min}\sim2$ kpc, confirming the
result from the previous analysis. As an example, we show in
Fig. \ref{fig:114kburn} these fits performed on the source 5C4.114,
where the effect of depolarization is more appreciable, thanks to the
20 cm observations.

%\begin{figure*}[htb]
%\includegraphics[width=0.45\textwidth]{./CHI2_soloRMScon.eps}
%\includegraphics[width=0.45\textwidth]{./best_profile_tot.ps}
%\caption{Left: $\chi^2$ plane obtained by comparing simulated and observed
%  $\sigma_{RM}$. Right: $\sigma_{RM}$ and $\langle RM \rangle$ for the best model
%  (cyan continuous line) and its dispersion (cyan dotted lines), given by the rms
%  of the different random realizations. Observed points are shown in
%  red. }
%\label{fig:chi2plane}
%\end{figure*}

\subsection{The magnetic field profile}
\label{sec:Bprof_sim}
\begin{figure*}[ht]
\includegraphics[width=\textwidth]{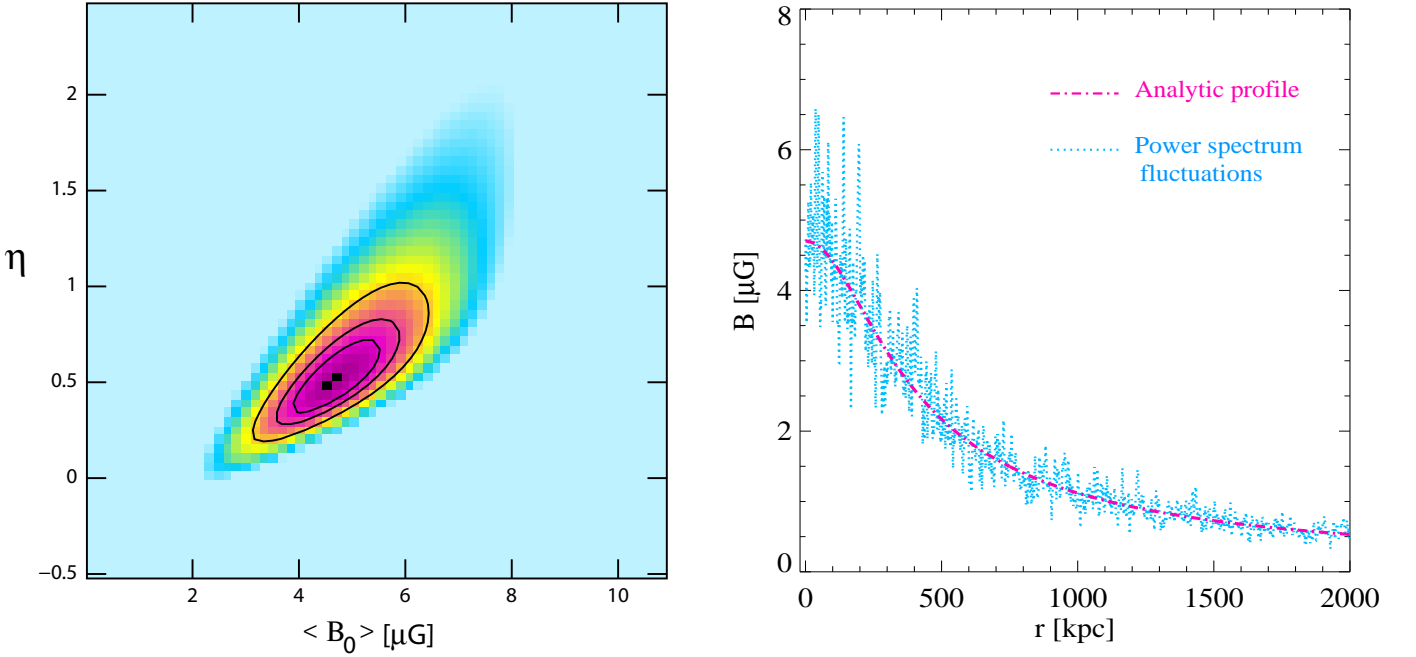}
\caption{{\it Left}: $\chi^2$ plane obtained by comparing simulated
  and observed $\sigma_{RM}$.{\it Right}: Profile of the best magnetic
  field model. magenta line refers to the analytic profile
  (Eq. \ref{eq:BProfile}), while the blue line refers to a slice
  extracted from the simulated magnetic field numerical model. Power
  spectrum fluctuations on the profile are shown.}
\label{fig:chi2plane}
\end{figure*}

The results obtained from the previous section indicate the power
spectrum that is able to best reproduce the observed RM images.  In
order to investigate the magnetic field radial profile we simulated 3D
Kolmogorov power spectra, with $\Lambda_{max}=$ 34 kpc and
$\Lambda_{min}=$ 2 kpc, as derived from the 2D analysis
(Sec. \ref{sec:PWsim}).  A computational grid of 2048$^3$ pixels was
used, and the pixel-size was fixed to 0.5 kpc. This guarantees that
the Nyquist criteria is satisfied for $\Lambda_{min}$ and that
fluctuations on scales $\sim \Lambda_{max}$ are also well represented
in the cube.  For each of these simulations Eq. \ref{eq:RM} was
integrated numerically, with a step of 0.5 kpc along the line of
sight.  The limits of the integral in Eq. \ref{eq:RM} were $[0; 10
  r_c]$ for the cluster's sources 5C4.85 and 5C4.81 and $[-10 r_c; 10
  r_c]$ for the other sources in the background of the cluster.  The
simulated field of view covers an area of 1024$\times$1024 kpc$^2$,
thus the cube has been replicated to achieve a field of view that is
large enough to reach the farthest source (5C4.152).\\ The integration
was repeated by varying the parameter $B_0$ in the range [0.1; 11]
$\mu$G, with a step of $\sim$ 0.17 $\mu$G, and $\eta$ in the
range[-0.2; 2.5] with a step of 0.04. For each combination of $B_0$
and $\eta$ a RM simulated image was thus obtained covering the full
cluster area.\\ We extracted from this RM image seven fields, each
lying in the plane of the sky in the same position of the observed
sources, and having the same size of the observed RM images. The
simulated RM images were convolved with a Gaussian beam having
FWHM$=$0.7$\times$0.7 kpc, in order to have the same resolution of the
observations. Finally the simulated RM fields were blanked in the same
way as the corresponding RM source.\\ The result of this integration
is, for each combination of ($B_{0}$;$\eta$), a set of seven simulated
RM images, that are subject to the same statistical biases of the
observed images.\\ This process was repeated 50 times, each starting
from a different random seed to generate the magnetic field power
spectrum model.\\ For each source and for each value of
($B_{0}$;$\eta$) a simulated RM image was obtained for every
realization of the same power spectrum model. The mean and the
standard deviation of the $\sigma_{RM,sim}(B_0,\eta)$ was computed
from the simulated RM images, and then the $\chi^2$ was obtained
(Eq. \ref{eq:chi2}).  The resulting $\chi^2$ plane is shown in
Fig. \ref{fig:chi2plane}.  The minimum value is achieved for
$B_{0}=$4.7 $\mu$G and $\eta=0.5$, but the 1-$\sigma$ confidence level
of the $\chi^2$ indicates that values going from $B_{0}=$3.9 $\mu$G and
$\eta=0.4$, to $B_{0}=$5.4 $\mu$G and $\eta=0.7$, are equally
representative of the magnetic field profile, according to the
degeneracy between the two parameters. Magnetic field models with a
profile flatter than $\eta < 0.2$ and steeper than $\eta > 1.0$ are
excluded at 99\% confidence level, for any value of $\langle B_0
\rangle$. Also magnetic field models with $\langle B_0 \rangle < 3.1
\mu$G and $\langle B_0 \rangle >$ 6.5 $\mu$G are excluded at the 99\%
confidence level for any value of $\eta$.  It is interesting to note
that the best models include $\eta=0.5$, the value expected in the
case of a magnetic field energy density decreasing in proportion to
the gas energy density (assuming a constant average gas temperature),
and $\eta=0.67$, expected in the case of a magnetic field frozen into
the gas. In the latter case the corresponding value of $B_0$ is
$\sim$5.2 $\mu$G. \\ 

The knowledge of the magnetic field strength and
structure in the ICM has strong implications for models explaining the
formation of diffuse radio sources like radio halos. Testing the
different models proposed in the literature is beyond the scope of
this paper. We point out, however, that cosmological simulations
recently performed by Donnert et al. (2009b) have shown that it is
possible to test a class of these models once the magnetic field
profile is known.\\
\subsection{Results excluding the source 5C4.74}
The same procedure described above has been repeated excluding the
source 5C4.74 (see Sec. \ref{sec:74}). The minimum value for the
$\chi^2$ is again achieved with a model characterized by $B_0=$4.7
$\mu$G and $\eta=0.5$. This is not surprising since the fit is
computed based on $\sigma_{RM}$, and the source is sampled with only
10 beams. In order to investigate possible effects arising from
  the interaction of the Coma cluster with the NGC4839 group RM images
  of more sources in this region would be required.

%---------------------------------------------------------------------

\section{Comparison with other estimates}
\label{sec:comparison}
In the literature there is a long-standing debate on the magnetic
field strength derived from the RM analysis compared to the
equipartition estimate and to the Inverse Compton hard X-ray
emission. The discrepancy may arise from the different (but not
incompatible) assumptions, and, moreover, are sensitive to the
magnetic field on different spatial scales. Assuming the magnetic
field models derived in the previous section, it is possible to derive
an estimate that is comparable with equipartition values, and with the
Inverse-Compton detection as well as with the upper limits derived
from new hard X-ray observations.  In order to obtain a value that is
directly comparable with the equipartition magnetic field estimate, we
have to derive the average magnetic field strength resulting from our
RM analysis over the same volume assumed in the equipartition
analysis, that is $\sim$1 Mpc$^3$. The magnetic field model resulting
from our RM analysis gives an average magnetic field strength of
$\sim$ 2 $\mu$G, consistent with the equipartition estimate derived
from the radio halo emission ( 0.7 - 1.9 $\mu$G Thierbach et
al. 2003), despite the different assumptions that these two methods
require.\\ The Inverse Compton hard X-ray emission has been observed
with the {\it Beppo Sax} satellite. Its field of view is
$\sim1.3^{\circ}$, corresponding to $\sim 2.2 \times 2.2$ Mpc$^2$ at
the Coma redshift.
We computed the average value of the magnetic field over the same
volume sampled by {\it Beppo Sax}. We obtained $\sim$0.75 $\mu$G when
the best model is assumed, that is a factor four higher than the value
derived from Hard-X ray observations (Fusco Femiano et al. 2004). We
note however that models compatible with our data within 1-$\sigma$
of the $\chi^2$ give values slightly different, going from 0.9 to 0.5
$\mu$G. The steepest magnetic field model that is compatible with our
data at 99\% confidence level ($B_{0}\sim 6.4$ $\mu$G, $\eta=0.95$)
gives 0.2 $\mu$G when averaged over the volume corresponding to the
{\it Beppo Sax} field of view. Deeper Hard-X ray observations would be
required to better compare the two estimates. The values computed here
indicate however that they can be reconciled.  Recently, new hard
X-ray observations of the Coma cluster have been performed with the
new generation of satellites (see the work by Wik et al. 2009 using
XMM and Suzaku data, Lutovinov et al. 2008 using ROSAT, RXTE and
INTEGRAL data, Ajello et al. 2009 using XMM-Newton, Swift/XRT, Chandra
and BAT data). These observations failed to find statistically
significant evidence for non-thermal emission in the hard X-ray
spectrum of the ICM, which is better described by a single or
multi-temperature model.  Given the large angular size of the Coma
cluster, if the non-thermal hard X-ray emission is more spatially
extended than the observed radio halo, both Suzaku HXD-PIN and BAT
Swift may miss some fraction of the emission. These efforts have thus
derived lower limits for the magnetic field strength, over areas
smaller than the radio halo. The lower limit reported by Wik et
al. (2009) is e.g.  $\langle B \rangle >$ 0.2 $\mu$G, that is
compatible with our results.

\begin{figure}[h]
\includegraphics[width=0.45\textwidth]{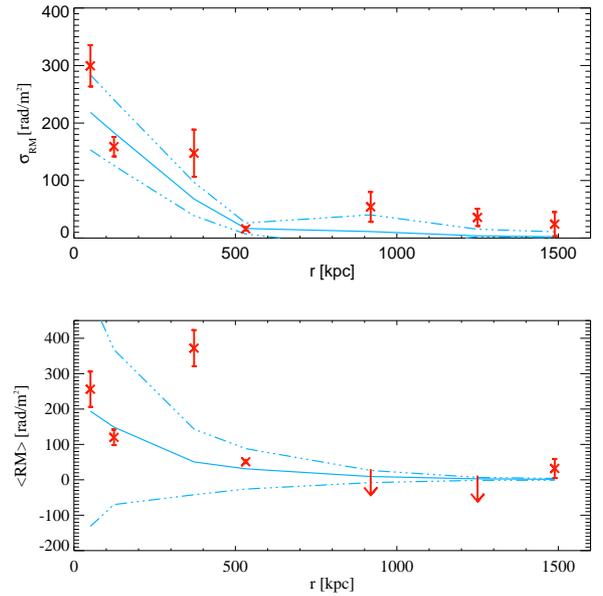}
\caption{$\sigma_{RM}$ and $\langle RM \rangle$ for the best model
  (cyan continuous line) and its dispersion (cyan dotted lines), given
  by the rms of the different random realizations. Observed points are
  shown in red. }
\label{fig:bestfit}
\end{figure}

%---------------------------------------------------------------------
\section{Conclusions}
\label{sec:concl}
We have presented new VLA observations of seven sources in the Coma
cluster field at multiple frequencies in the range 1.365 -- 8.465
GHz. The high resolution of these observations has allowed us to
obtained detailed RM images with 0.7 kpc resolution. The sources were
chosen in order to sample different lines-of-sight in the Coma cluster
in order to constrain the magnetic field profile. 
We used the numerical approach
proposed by Murgia et al. (2004) to realize 3D magnetic
field models with different central intensities and radial slopes, and
derived several realizations of the same magnetic field model in order
to account for any possible effect deriving from the random nature of
the magnetic field. Simulated RM images were obtained, and
observational biases such as noise, beam convolution and limited
sampled regions were all considered in comparing models with the data.
We found that $\sigma_{RM}$ and $\langle RM \rangle$ decrease with
increasing distance from the cluster center, except for the source
5C4.74, that shows a high value of $\langle RM \rangle$. We argue that
this may arise from its peculiar position southwest of the Coma cluster
core, toward the NGC4839 group that is currently merging with the Coma
cluster.\\ Our results can be summarized as follows:
\begin{itemize}
\item{the RM ratio and the DP ratio were used to analyze the magnetic
  field power spectrum. Once a Kolmogorov index is assumed, the
  structure-function, the auto-correlation function and the
  multi-scale statistic of the RM images are best reproduced by a
  model with $\Lambda_{max}=$ 34 kpc and $\Lambda_{min}=$ 2 kpc. We
  performed a further check to investigate the best value of
  $\Lambda_{min}$ by fitting the Burn law (Burn 1966).  This confirmed
  the result obtained from the previous analysis.}
\item{The magnetic field radial profile was investigated through a
  series of 3D simulations. By comparing the observed and simulated
  $\sigma_{RM}$ values we find that the best models are in the range
  ($B_0=$3.9 $\mu$G;$\eta=$0.4) and ($B_0=$5.4 $\mu$G;$\eta=$0.7).  It is
  interesting to note that the values $\eta=$0.5 and 0.67 lie in this
  range. They correspond to models where the magnetic field energy
  density scales as the gas energy density, or the magnetic field is
  frozen into the gas, respectively. This is expected from a
  theoretical point-of-view since the energy in the magnetic component
  of the intracluster medium is a tiny fraction of the thermal
  energy. Values of $B_{0}>$7 $\mu$G and $<$3 $\mu$G as well as $\eta
  < 0.2$ and $\eta > 1.0$ are incompatible with RM data at the 99\%
  confidence level. }
\item{The average magnetic field intensity over a volume of $\sim$ 1
  Mpc$^3$ is $\sim$ 2 $\mu$G, and can be compared with the
  equipartition estimate derived from the radio halo
  emission. Although based on different assumptions, and although the
  many uncertainties relying under the equipartition estimate, the
  model derived from RM analysis gives an average estimate that is
  compatible with the equipartition estimate. A direct comparison with
  the magnetic field estimate derived from the IC emission is more
  difficult, since the Hard-X detection is debated, and depending on
  the particle energy spectrum, the region over which the IC emission
  arises may change. The model derived from RM analysis gives a
  magnetic field estimate that is consistent with the present lower
  limits obtained from hard X-ray observations. The values we obtain
  for our best models are still a bit higher when compared with the
  estimate given by Fusco Femiano et al. (2004). It is worth to
  remind, as noted by several authors (see Sec. \ref{sec:comparison}),
  that the IC estimate derived from Hard X-ray observations could be
  dominated by the outer part of the cluster volume, where the
  magnetic field intensity is lower, depending on the spatial and
  energy distribution of the emitting particles. Future Hard-X ray
  missions could help in clarifying this issue.}
\end{itemize}
 
\bigskip
{\bf Acknowledgements} A.B. is grateful to the people at the
Osservatorio Astronomico di Cagliari for their kind hospitality. We
thank R. Fusco Femiano and G. Brunetti for useful discussions. This
work is part of the ``Cybersar'' Project, which is managed by the
COSMOLAB Regional Consortium with the financial support of the Italian
Ministry of University and Research (MUR), in the context of the
''Piano Operativo Nazionale Ricerca Scientifica, Sviluppo Tecnologico,
Alta Formazione (PON 2000-2006)''. K.~D.~acknowledges the supported by
the DFG Priority Programme 117. NRAO is a facility of the National
Science Fundation, operated under cooperative agreement by Associated
Universities. This work was partly supported by the Italian Space
Agency (ASI), and by the Italian Ministry for University and research
(MIUR). This research has made use of the NASA/IPAC Extragalactic Data
Base (NED) which is operated by the JPL, California institute of
Technology, under contract with the National Aeronautics and Space
Administration.

\appendix
\begin{onecolumn}
\section{Structure Function and Multi-Scale Statistics with different power spectrum models}
\label{sec:app}
In this Appendix we discuss how other power-law spectral models could
be representative of the data presented in the paper.  The analysis is
performed on the basis of the the structure-function, auto-correlation
function and multi-scale statistics(MSS). Following the approach
discussed in Sec. \ref{sec:PWsim}, we have obtained simulated RM
images from different power spectrum models and compared them with
observed data.  We show in Fig. \ref{fig:AppK} the structure function,
auto-correlation function and MSS for a Kolmogorov power spectra that
has different $\Lambda_{max}$.  We show in Fig.  \ref{fig:AppN2} the
fit to the structure, auto-correlation functions and MSS for power
spectrum models with $n=2$, and different values of
$\Lambda_{max}$. We show only the plots obtained for the central
source 5C4.85. These figures demonstrate how the RM data presented in
this paper are sensitive to different power spectrum models.\\ We note
that Kolmogorov power spectra with $\Lambda_{max}\sim$100 and 10 kpc
fail in reproducing the $\langle RM \rangle$. These trends can be
easily understood since power spectrum models with $n>3$ have most of
the magnetic energy on large spatial scales, and thus small changes in
$\Lambda_{max}$ have a consistent impact on the resulting
statistics. According to results presented in Sec. \ref{sec:lambdan},
the case $\Lambda_{max}=$20 kpc gives a reasonable fit to our
data. The best fit is however achieved for $\Lambda_{max}=$34 kpc.  In
Fig. \ref{fig:AppN2} similar fits obtained for power spectra models
with $n=2$ are shown. As indicated by the analysis performed in
Sec. \ref{sec:lambdan}, in this case the best agreement with
observations is achieved for $\Lambda_{max}$ of order of hundreds
kpc. We note that because of the power spectrum degeneracy, it is
possible to obtain a reasonable fit to our data. Indeed the case
$\Lambda_{max}=$400-800 kpc can reproduce the MSS statistics, although they
fail in reproducing the $S(r)$ trend on large spatial scales,
indicating that a larger value of $n$ is required. 
\begin{figure*}[h] 
\centering
\includegraphics[width=\textwidth]{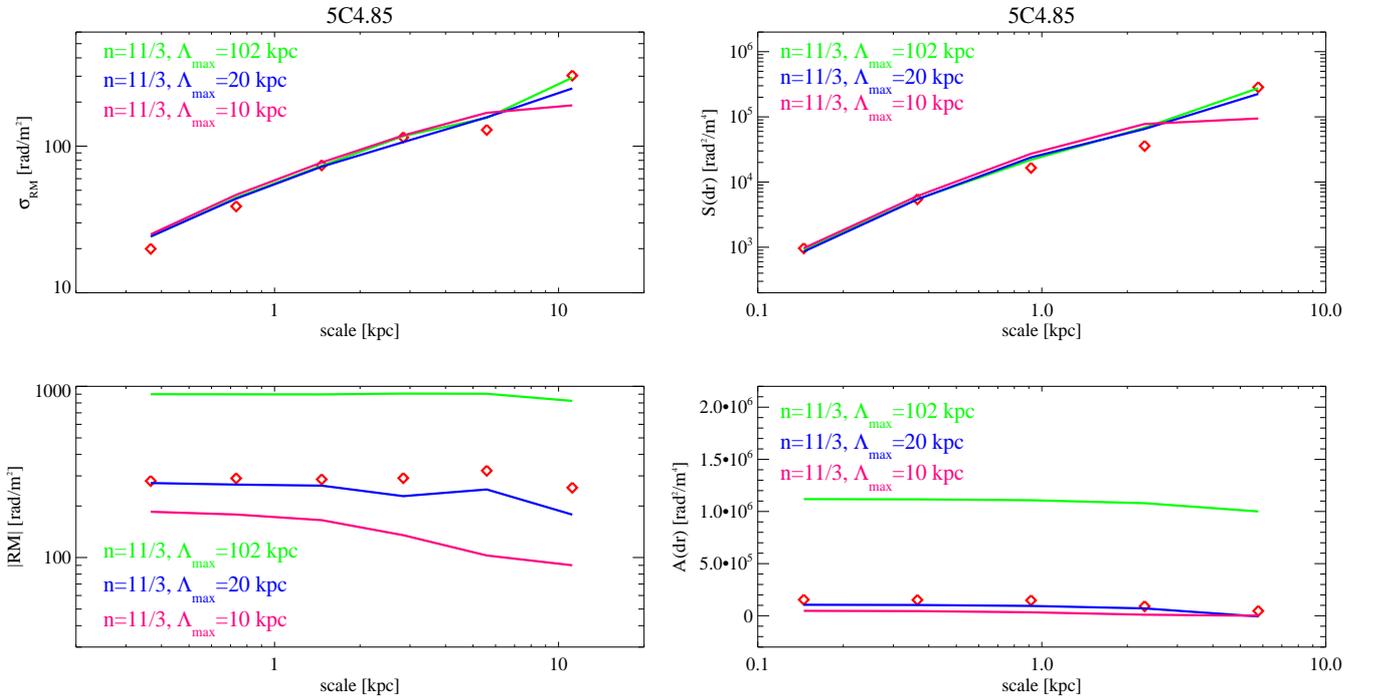}
\caption{Fit to the RM images for different Kolmogorov power spectra
  for the central sources 5C4.85. The different models are indicated
  by different colors (see labels) {\it left}: fit to the $\sigma_{RM}$
  and $\langle RM \rangle$; {\it right}: fit to the $S(r)$ and
  $A(r)$. Red diamonds represent the observed statistics. Lines
  represent the value obtained by averaging 10 power spectra generated
  with different random seeds.}
\label{fig:AppK}
\end{figure*}

\begin{figure*}[ht] 
\centering
\includegraphics[width=\textwidth]{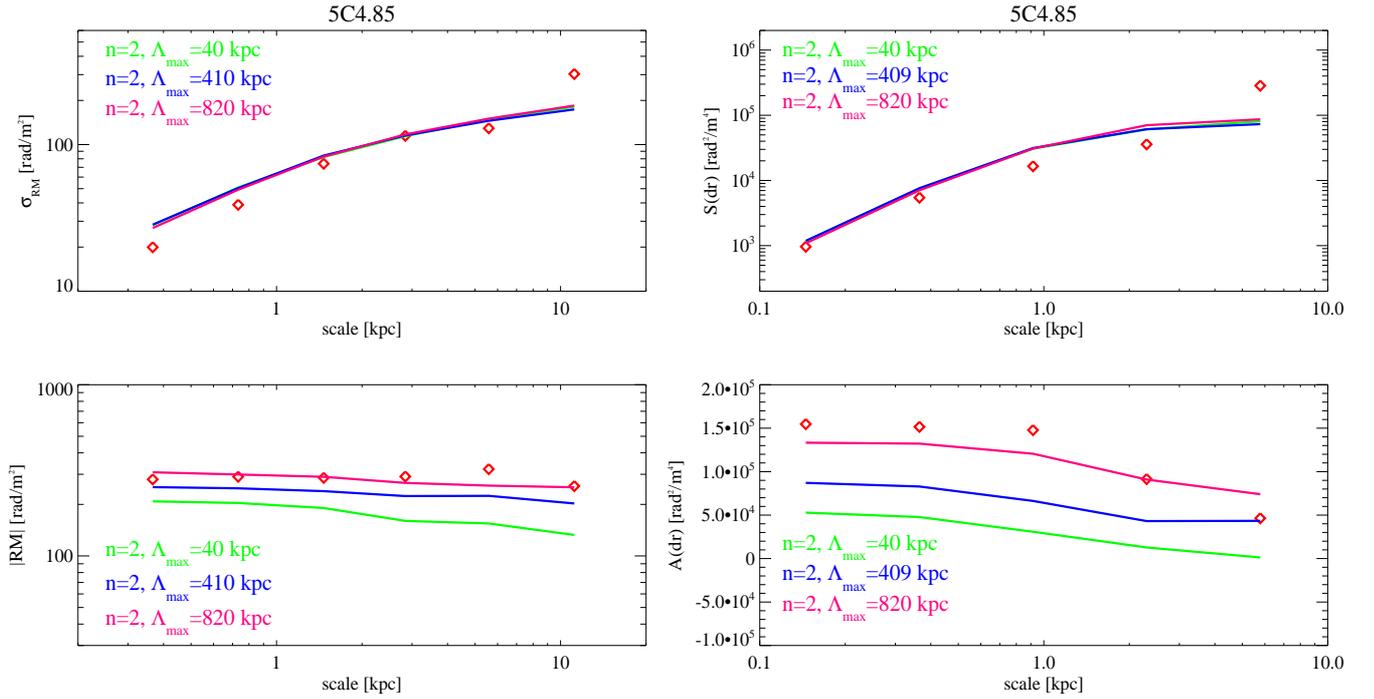}
\caption{Fit to the RM images for different  power spectra with $n=2$
  for the central sources 5C4.85. The different models are indicated
  by different colors (see labels) {\it left}: fit to the $\sigma_{RM}$
  and $\langle RM \rangle$; {\it right:} fit to the $S(r)$ and
  $A(r)$. Red diamonds represent the observed statistics. Lines
  represent the value obtained by averaging 10 power spectra generated
  with different random seeds. }
\label{fig:AppN2}
\end{figure*}

\section{Limits on the magnetic field profile from background radio sources.}
Although several arguments (see Sec. \ref{sec:RMlocal}) suggest
  that the main contribution to the observed RMs is due to the ICM,
  the best way to firmly avoid any kind of local contribution would be
  to consider only background radio galaxies in the analysis. This is
  however not trivial in general and not feasible here. In fact,
  sources located in the inner region of the cluster, at distances
  $\leq (1-2) r_C$ are fundamental to constrain the magnetic field
  strength and radial decline, so that a peculiar cluster where
  background bright and wide sources are seen in projection very close
  to the cluster center would be required. These conditions are not
  fulfilled in the case of the Coma cluster, even though it is a very
  nearby cluster, where several lines of sight can be inspected. We
  show in Fig. \ref{fig:appProfile} (left panel) the $\chi^2$ plane
  obtained by considering only the background radio galaxies: 5C4.74,
  5C4.114, 5C4.127, 5C4.42 and 5C4.152.  This plot shows that for
  every value of $B_0$ it is possible to find a value of $\eta$ that
  can reproduce the observed data within 1-$\sigma$ confidence
  level. In the same Fig. in the right panel we show the trends of
  $\sigma_{RM}$ and $\langle RM \rangle$ as a function of $r$ obtained
  for different value of $B_0$. They all lie within 1$\sigma$
  confidence level of the $\chi^2$ plane. It is clear from this plot
  that the missing information at projected distances $r <$ 300 kpc
  does not permit us to infer the magnetic field strength and radial
  decline in the Coma cluster. Even unrealistic models where
  $\eta\leq$0 cannot be ruled out when the two wide central sources
  are not considered. Future instruments such as SKA are expected to
  detect many more radio sources seen through a single cluster,
  possibly allowing this kind of analysis to be performed excluding
  cluster members.
\begin{figure*}[ht] 
\centering
\includegraphics[width=\textwidth]{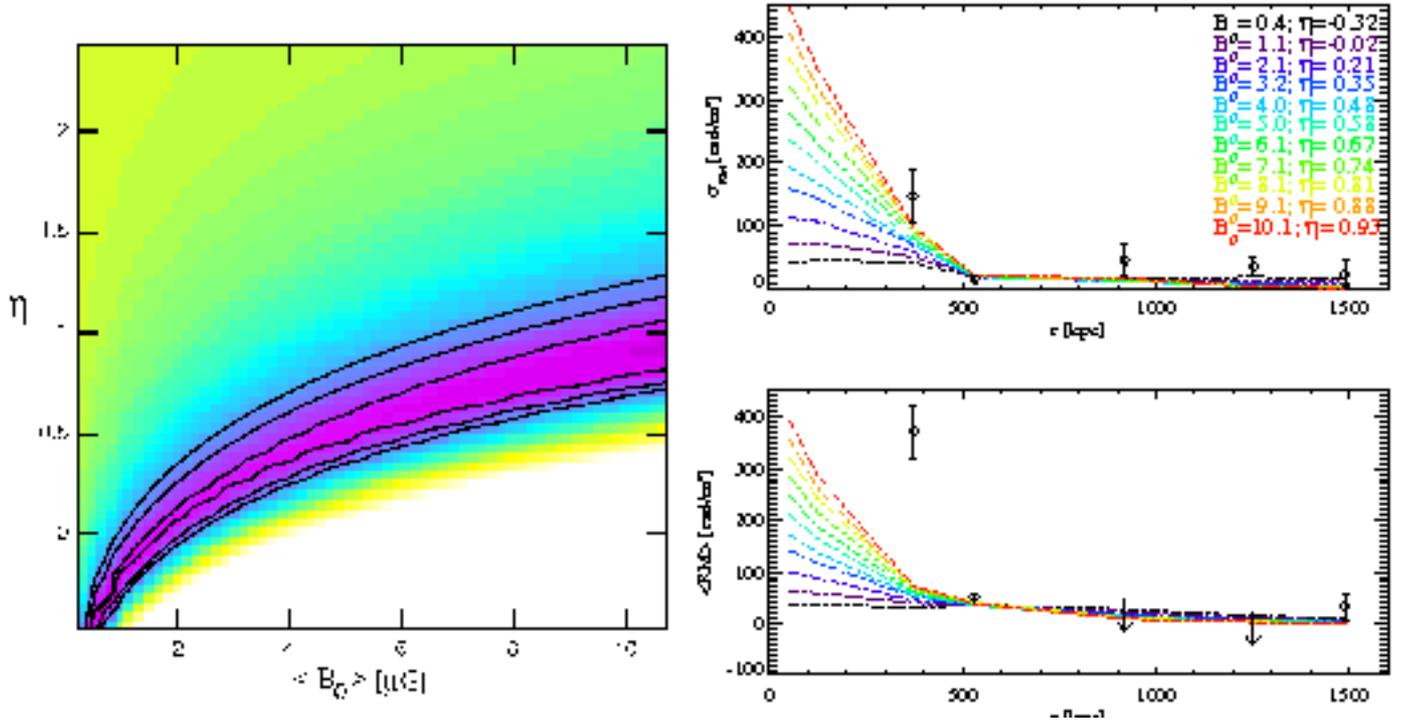}
\caption{{\it Left:} $\chi^2$ plane obtained by comparing simulated
  and observed $\sigma_{RM}$ for background sources. Lines refer to 1,2
  and 3-$\sigma$ confidence level. {\it Right:} $\sigma_{RM}$ and
  $\langle RM \rangle$ trends for different models that lie within
  1-$\sigma$ confidence level of the $\chi^2$.}
\label{fig:appProfile}
\end{figure*}

\end{onecolumn}

\end{document}